\documentclass[journal=ancac3,manuscript=article]{achemso}

\usepackage{amsmath}
\usepackage{chemformula}
\usepackage{chemist}
\usepackage{color}
\usepackage{graphicx}
\usepackage[colorlinks=true,
            linkcolor=blue,
            citecolor=blue,
            filecolor=blue,
            urlcolor=blue
           ]{hyperref}
\usepackage[version=4]{mhchem}
\usepackage{soul}
\usepackage{units}
\usepackage{xcolor}

\usepackage{array}
\usepackage{booktabs}
\usepackage{multicol}
\usepackage{multirow}
\newcolumntype{C}[1]{>{\centering\arraybackslash}p{#1}}

\newcommand{\phys}{
    Department of Physics,
    Chalmers University of Technology,
    412~96 G\"{o}teborg, Sweden
}

\newcommand{\warsaw}{
    Faculty of Physics,
    University of Warsaw,
    Pasteura 5,
    02-093 Warsaw, Poland
}
\newcommand{\linkoping}{
    Department of Physics,
    Link\"{o}ping University,
    581 83 Link\"{o}ping, Sweden
}

\author{Abhay V. Agrawal}
\affiliation{\phys}
\author{Alexander Yu. Polyakov}
\affiliation{\phys}
\author{Jens Eriksson}
\affiliation{\linkoping}
\author{Tomasz J. Antosiewicz}
\affiliation{\warsaw}
\alsoaffiliation{\phys}
\author{Timur O. Shegai}
\affiliation{\phys}
\email{timurs@chalmers.se}

\title{Humidity-enhanced \ce{NO2} gas sensing using atomically sharp edges in multilayer \ce{MoS2}}

\begin{document}



\begin{abstract}

  Ambient humidity poses a significant challenge in the development of practical room temperature \ce{NO2} gas sensors. Here, we employ atomically precise zigzag edges in multilayer \ce{MoS2}, fabricated using electron beam lithography and anisotropic wet etching, to achieve highly sensitive and selective gas sensing performance that is humidity-tolerant at elevated temperatures and humidity-enhanced at room temperature under ultraviolet illumination. Notably, exposure to 2.5 parts per billion (ppb) \ce{NO2} at 70\% relative humidity under ultraviolet illumination and at room temperature resulted in a 33-fold increase in response and a 6-fold faster recovery compared to 0\% relative humidity, leading to response values exceeding 1100\%. The optimized samples demonstrated a theoretical detection limit ranging from 4 to 400 parts per trillion (ppt) \ce{NO2}. The enhanced \ce{NO2} sensing capabilities of \ce{MoS2} edges have been further confirmed through first-principles calculations. Our study expands the applications of nanostructured \ce{MoS2} and highlights its potential for detecting \ce{NO2} at sub-ppb levels in complex scenarios, such as high humidity conditions.

\end{abstract}





\section{Introduction}
Detecting \ce{NO2} is crucial in modern society, as it is a common industrial byproduct of any high-temperature combustion
in nitrogen containing environments~\cite{atkinson2000atmospheric, vineis2004outdoor}. \ce{NO2} interacts with water and volatile organic compounds (VOCs) to contribute to the formation of acid rain and photochemical smog. Prolonged exposure to even low concentrations, such as 1 part per million (ppm) of \ce{NO2}, can lead to the development of chronic bronchitis, respiratory-related diseases, and ocular discomfort~\cite{vineis2004outdoor, schwela2000air}. Furthermore, \ce{NO2} is extensively used in the production of nitric acid, a substance that finds a widespread use~\cite{hodge1994pollution}. A pulmonary condition, known as asthma, is often associated with elevated levels of nitric oxides (\ce{NO2} and NO) in the exhaled breath. While the exhaled nitric oxide (NO) concentration typically ranges from 20 to 25 parts per billion (ppb) in healthy individuals, those with asthma exhibit elevated levels ranging from 35 to 50 ppb, regardless of age~\cite{korn2020measurement}. Recognizing these health risks, the United States Environmental Protection Agency (U.S. EPA) has set a regulated limit for exposure to \ce{NO2} at 53 ppb~\cite{esworthy2013national}. Consequently, there is a pressing need for reliable sensors capable of detecting minute concentrations of \ce{NO2} below the regulatory threshold~\cite{esworthy2013national, guarnieri2014outdoor}.
 
Various methods have been employed to detect \ce{NO2} at ppb levels using \ce{MoS2}. \ce{MoS2} is a semiconducting material exhibiting a thickness-dependent bandgap (1.2-1.8 eV), high surface-to-volume ratio, a wide spectrum of light absorption from ultraviolet (UV) to near-infrared (near-IR), as well as excellent catalytic and gas sensing properties~\cite{splendiani2010emerging, radisavljevic2011single, jaramillo2007identification}. Edges of \ce{MoS2} have attracted research interest as favorable sites for gas adsorption, surpassing their basal plane counterparts. This phenomenon stems from the existence of vacancies (and other defects) supporting high binding energies and charge transfer values for target gas molecules~\cite{yue2013adsorption, cho2015highly, agrawal2018photoactivated}. It has been observed both theoretically and experimentally that the adsorption of \ce{NO2} molecules is favorable at \ce{MoS2} edges in comparison to other gases such as \ce{H2}, CO, \ce{NH3}, \ce{H2S}, \ce{CO2}, and \ce{CH4}~\cite{yue2013adsorption, cho2015highly, agrawal2018photoactivated}. Various advanced methods for improved gas sensing using \ce{MoS2} include decoration of the latter with metal nanoparticles (NPs), fabricating van der Waals heterostructures, and morphology engineering. For example, Kim \textit{et al.} functionalized a solution-based \ce{MoS2} surface with different noble metal NPs~\cite{kim2023drastic}. The selective character of \ce{MoS2} sensors may be modified by doping with metal NPs in order to target a specific gas. Long \textit{et al.} proposed a \ce{MoS2}/graphene heterostructure for ultrasensitive \ce{NO2} detection. At room temperature (RT), this hybrid structure demonstrated a 50 ppb \ce{NO2} detection~\cite{long2016high}. A notable advantage of \ce{MoS2} heterostructures is their rapid charge separation, which has the potential to develop gas sensors with fast response and recovery times. Although these strategies could pave the way to developing ultra-sensitive and selective \ce{NO2} sensors, both methods introduce fabrication complexity in device manufacturing. Furthermore, the use of noble metal NPs is potentially costly.

Recently, gas sensors based on edge-enriched \ce{MoS2} have emerged as a promising prospect. Edge-enriched \ce{MoS2} offers a higher surface area and more active edge sites. Real-world devices and applications such as transistors, catalysts, photodetectors, and gas sensors based on nanopatterned transition metal dichalcogenides (TMDs), especially \ce{MoS2}, can experience significant improvements in their properties through edge enrichment~\cite{ShimNanoporousMoS2Field-Effect, Arindam, Anamika, park2021nano, munkhbat2023nanostructured, cho2015highly}. Cho \textit{et al.} compared gas adsorption on three distinct morphologies of \ce{MoS2} (horizontally aligned (HA) \ce{MoS2}, vertically aligned (VA) \ce{MoS2} and a mixture of HA and VA \ce{MoS2}) grown by the combination of radio frequency (RF) sputtering and rapid sulfurization~\cite{cho2015highly}. The authors showed theoretically that zigzag edges in \ce{MoS2} provide an 18 times higher binding energy for \ce{NO2} molecules adsorption compared to the basal plane. The experimental and theoretical investigations indicated that edge-enriched \ce{MoS2} has a superior sensitivity for \ce{NO2} gas molecules. Agrawal \textit{et al.} proposed chemical vapor deposition (CVD) grown mixed in-plane and edge-enriched \ce{MoS2} flakes~\cite{agrawal2018photoactivated}. The proposed structures showed a selective and sensitive nature for \ce{NO2} sensing at room temperature. Li \textit{et al.} used the hydrothermal method to prepare hollow, solid, and smooth \ce{MoS2} nanospheres. The sensing area for \ce{NO2} was doubled, and the response was enhanced 3.1 times in hollow \ce{MoS2} in comparison to smooth nanospheres. Edges in \ce{MoS2} can be either zigzag or armchair, among which the zigzag edges are thermodynamically more stable and most promising for gas molecule adsorption~\cite{xiao2016edge}. However, in most morphology-driven \ce{NO2} sensors, it is challenging to accurately control the precise edge orientation.

The top-down nanopatterning can substantially modify the structural, optical, and electrical characteristics TMDs materials. These properties are strongly influenced by changes in the atomic-scale structure and defects~\cite{munkhbat2020transition, jessen2019lithographic, munkhbat2023nanostructured}. Numerous methods, such as atomic force lithography, plasma treatment, block copolymer lithography (BCP), thermal annealing, laser writing, and seed-assisted growth, have been used previously~\cite{liu2023patterning, cao2013direct, enrico2023ultrafast, chang2023manipulating, way2018seed, yu2011control}. However, these approaches lead to reduced accuracy and sharpness, random orientation, uneven patterning, complex fabrication stages, and limited control over the size and shape of the nanostructures. In seed-assisted patterning, for instance, the resulting TMD materials may exhibit random orientation and face a continuous risk of contamination from seeds throughout the growth process~\cite{way2018seed, li2018site}. The laser writing and thermal annealing methods potentially eliminate the necessity for masks and photoresists, simplifying the nanopattern design~\cite{ryu2017sub, liu2020thermomechanical}. Similarly, BCP lithography has emerged as a promising strategy for nanopatterning. However, in all these processes, the patterns are typically randomly orientated, not precisely controlled, disordered, and contain structural defects~\cite{park2021nano}.

In addition to the aforementioned challenges, it is important to recognize that the relative humidity (RH) commonly present in the atmosphere poses a significant obstacle to the development of real-time gas sensors using semiconducting materials. The water concentration in ambient air (atmospheric pressure, 25$^\circ$C) is 6,280 ppm at 20\% RH and 25,740 ppm at 80\% RH levels, respectively~\cite{yoon2016new}. Moreover, in real-life scenarios, the humidity levels are strongly dependent on environmental conditions, such as wind, rainfall, temperature, as well as day-night and seasonal oscillations. At room temperature, water adsorption leads to significant fluctuations in baseline resistance with variations in humidity~\cite{reddeppa2020nox, wang2019synergistic, chen2021uv}. Additionally, exhaled human breath contains substantial RH levels, which presents a challenge for reliable chemiresistive gas sensor operation. A commonly employed approach to overcome these challenges is elevated temperatures. Furthermore, various approaches to develop humidity-tolerant chemiresistive gas sensors, such as doping with metal NPs (Pd, Pt, Ag, Au), element doping (Ce, Pr, and Tb), integration with hydrophobic materials (polydimethylsiloxane (PDMS), prepared polyaniline (PANI), graphite nanoflakes) \textit{etc.} have been previously employed~\cite{ma2015pd, wang2012templating, yang2020acetone, qin2021dual, yoon2016new, kim2019humidity, li2020hydrophobic, maity2018polyaniline}. Although such methods improved the anti-humidity ability of the respective gas sensors, each mentioned approach added complexity, hindering their practical use. 
 
Here, we report sub-ppb-level \ce{NO2} sensors based on nanopatterned multilayer \ce{MoS2} exposing precise atomically sharp zigzag edges. The nanopatterned \ce{MoS2} resulted in honeycomb nanomesh structures fabricated by optimizing the process reported earlier~\cite{munkhbat2020transition}. The method provides a uniform, precise, and controllable mesh of \ce{MoS2} nanoribbons with widths ranging from a few hundred nanometers (nm) to sub ten nm (see Figure~\ref{fig:1}b-d). We studied the concentrations of \ce{NO2} gas in the range of 2.5 -- 10 ppb, below the threshold limit set by the U.S. EPA. The density of exposed edges is proportional to the hexagonal holes' density and their dimensions. We systematically studied the density-dependent \ce{NO2} sensing in several honeycomb \ce{MoS2} nanomesh samples. The \ce{NO2} response was best at 200$^\circ$C in the optimized honeycomb \ce{MoS2} nanomesh structures due to numerous exposed, uniform, precise zigzag edges, combined with optimal electrical performance. The studied samples demonstrated selective \ce{NO2} sensing (against CO, \ce{CH4}, \ce{H2}, and \ce{C2H6}) and demonstrated an extraordinary limit of detection ranging from 4 to 400 ppt, depending on the exact mode of operation. In addition, we investigated the room temperature operation of our sensors under high RH levels (up to 70\%) with and without UV illumination. Surprisingly, at RT the sensors showed a significantly enhanced response to \ce{NO2} in a high-humidity environment. Under UV illumination, this humidity-enhanced response exceeds 1100\% upon exposure to 2.5 ppb \ce{NO2}. Importantly, the humidity-enhanced sensing behavior is universal and was observed for other tested gases, including CO, \ce{CH4}, \ce{H2}, and \ce{C2H6}, albeit at much lower response levels than for \ce{NO2}. This universal humidity-enhanced behavior suggests that the mechanism of the enhanced sensor performance is also universal, and likely involves competition between \ce{H2O} and \ce{O2} adsorbed on the \ce{MoS2} surface under the influence of UV illumination. Our density functional theory (DFT) calculations confirm that \ce{NO2} adsorption is significantly influenced by the presence of edges and sulfur vacancies. Thus, the honeycomb \ce{MoS2} nanomesh samples, studied here, offer an optimal configuration, featuring a high density of sulfur vacancy sites along the multilayer hexagon rims. This unique structure enhances the \ce{NO2} adsorption capacity, making the honeycomb \ce{MoS2} nanomesh particularly effective for chemiresistive gas sensing applications in high humidity conditions.

\section{Results and discussion}

\subsection{Nanofabrication, morphology, and structure of honeycomb \ce{MoS2} nanomesh sensors}

\begin{figure}[ht!]
    \centering
    \includegraphics[width=0.99\linewidth]
    {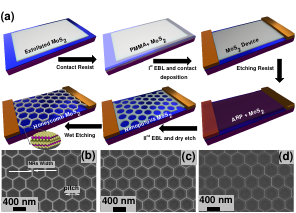}
    \caption{(a) Schematic illustration of the fabrication process towards honeycomb \ce{MoS2} nanomeshes. (b-d) The SEM images of deterministically nanofabricated, highly uniform, and controlled honeycomb nanomeshes with precise zigzag edges. The nanomeshes are formed by triangular arrays of holes with fixed pitch 400 nm and variable radius of the initial circular holes (b) 280 nm (c) 290 nm (d) 300 nm.}
    \label{fig:1}
\end{figure}

Figure~\ref{fig:1} illustrates the fabrication process for honeycomb \ce{MoS2} nanomeshes in multilayer \ce{MoS2}. A comprehensive exploration of the fabrication process is presented in the experimental section and elsewhere~\cite{munkhbat2020transition}. Using this method, we obtain hexagonal holes with atomically sharp zigzag edges. The scanning electron microscope (SEM) images of highly precise, uniform and large-area honeycomb \ce{MoS2} nanomeshes formed by numerous holes are shown in Figure~\ref{fig:1}b-d. The center-to-center distance (pitch) between the two neighboring circles (defined by electron beam lithography (EBL) and dry etching) was kept fixed at 400 nm while the circle's radius varied from 280 to 300 nm in steps of 10 nm. The nano-ribbons achieved by this unique method are down to sub-ten-nanometer width with ultrahigh precision, as depicted in Figure~\ref{fig:1}b-d.

\begin{figure}[ht!]
    \centering
    \includegraphics[width=0.99\linewidth]
    {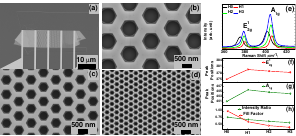}
    \caption{(a) The SEM image of tested devices. The high magnification SEM images of variable size hexagons (b) R = 450 nm and pitch -- 1450 nm. (c) R = 250 nm and pitch -- 820 nm. (d) R = 100 nm and pitch -- 325 nm. (e) The Raman spectra of various patterned devices. Variation in peak position as a function of various size hexagons (f) E$^{1}_{2g}$. (g) A$_{1g}$. (h) The change in Intensity ratio (E$^{1}$$_{2g}$/A$_{1g}$) with various size hexagons (blue). The calculated fill factor of each device (red)}
    \label{fig:2}
\end{figure}

A similar set of honeycomb \ce{MoS2} nanomeshes was prepared for the chemiresistive gas sensing assessment. The SEM images of the fabricated sensing devices are shown in Figure~\ref{fig:2}a-d. We designed the different testing devices on the same multilayer flake (30-40 nm thickness determined by profilometer) to minimize potential variations in electrical signal that may occur due to thickness, local defects, and fabrication processes and ensure a reliable comparison. In total, we fabricated four devices with channel lengths of 20 $\mu$m and widths of 13 $\mu$m, each. Among these four devices, three contain various honeycomb \ce{MoS2} nanomeshes, while one remains unpatterned. The nomenclature for the fabricated devices is as follows: the unpatterned device is labeled as H0, while the patterned devices are denoted as H1, H2, and H3. To test the density-dependent \ce{NO2} sensing, the initial size of the circles was varied. The density-dependent honeycomb meshes of H1, H2, and H3 devices were generated from the initial circular holes of 450, 250, and 100 nm radii, with nanomesh pitches -- 1450, 820, and 325 nm, respectively. The low-magnification SEM images of the devices are shown in Figure~\ref{fig:2}a-d and illustrate precise, clean, and ultra-sharp fabrication of honeycomb \ce{MoS2} nanomeshes with numerous zigzag edge sites available at the rims of hexagonal holes. Step-by-step optical images illustrating the nanofabrication process, starting from exfoliation and ending with honeycomb \ce{MoS2} nanopatterning are shown in the Supporting Information (SI), Figure S1. The SEM and optical images illustrate that the fabrication process does not damage \ce{MoS2} the remaining areas of the flake or leave photoresist residues on the sensing elements. The sensor chip mounted on the Pt100 heater with a Transistor Outline (TO8) header is shown in Figure S1e. The header with a sensor chip was installed inside our home-built sensing chamber and subsequently used for gas sensing experiments.

The precise zigzag nature of hexagon \ce{MoS2} was confirmed previously in detail using high-resolution transmission electron microscopy (HR-TEM)~\cite{munkhbat2020transition}. Here, we performed the Raman spectroscopy on patterned devices (H1, H2, and H3) and unpatterned device H0. Raman spectroscopy serves as a powerful and non-destructive technique, enabling the study of film orientation, doping, and the number of layers~\cite{iqbal2020review, chakraborty2012symmetry, kong2013synthesis}. Two distinct Raman modes, corresponding to the in-plane vibrations (E$^{1}_{2g}$) and out-of-plane vibrations (A$_{1g}$) of S and Mo atoms, have been observed, as depicted in Figure~\ref{fig:2}e. Both E$^{1}_{2g}$ and A$_{1g}$ peaks exhibit a blue shift upon nanopatterning compared to the unpatterned device, see Figure~\ref{fig:2}f-g. The A$_{1g}$ peak is sensitive to changes in electron concentration and the blue shift in A$_{1g}$ could signal a $p$-type doping in patterned devices~\cite{chakraborty2012symmetry,iqbal2020review}. A more comprehensive discussion of $p$-doping observed through electrical measurements is provided in subsequent sections. However, we also observe that both Raman peaks exhibit a slight redshift with increasing hexagon density (devices from H1 to H3). The redshift in the E$^{1}_{2g}$ and A$_{1g}$ modes can be attributed to terrace-terminated defects and an increase in \ce{MoS2} edge density~\cite{kong2013synthesis, kim2018structural}. The difference between the E$^{1}_{2g}$ and A$_{1g}$ peak positions, which is sensitive to the material thickness, is around 25 cm$^{-1}$ for all devices, implying the preservation of the device thicknesses with etching. E$^{1}_{2g}$ and A$_{1g}$ modes correspond to in-plane and out-of-plane vibrations, respectively. Thus, the peak intensity ratio of (E$^{1}_{2g}$/A$_{1g}$) can be used to identify the exposed edges in \ce{MoS2}~\cite{lee2010anomalous}. In our case, the peak intensity ratio is also decreased from 0.78 to 0.58 (in an H0 -- H3 row), indicating a clear increase in the number of edges with nanopatterning.


To confirm the increase in \ce{MoS2} edge content, we calculated the fill factor based on SEM images. The fill factor is defined as the ratio between the effective \ce{MoS2} area of the device to the area of the unpatterned device. The calculated fill factor is 1, 0.62, 0.48, and 0.42 for devices H0, H1, H2, and H3, respectively. These values quantify the increase in the edge content with the increasing hexagon density, see Figure~\ref{fig:2}h. The Raman signal originating from A$_{1g}$ mode or from edges is increased simultaneously with the decrease in fill factor.

\subsection{\ce{NO2} gas sensing}

We now turn our attention to the gas-sensing performance of our devices. The \ce{NO2} sensing experiments were conducted using synthetic air background (80\% O$_{2}$ and 20\% N$_{2}$). We tested four different concentrations of \ce{NO2}, ranging from 2.5 to 10 ppb, at two different operating temperatures: 200$^\circ$C and RT. The gas sensing performance was evaluated at a fixed bias voltage, $V_{bias} = +1$ V, via two-probe resistance measurements for all four devices. The initial current-voltage ($I-V$) analysis revealed a notable trend: an increase in $p$-doping levels correlated with the rise in hexagon density within the fabricated devices. Detailed $I-V$ characteristics are provided in Figure S2. 

\subsubsection{\ce{NO2} sensing at 200$^\circ$C}

\begin{figure}[ht!]
    \centering
    \includegraphics[width=0.99\linewidth]
    {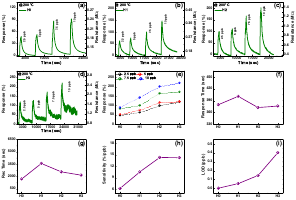}
    \caption{The dynamic \ce{NO2} sensing response (2.5 -- 10 ppb) at 200$^\circ$C and 0\% humidity obtained from (a) the unpatterned device (H0) and (b-d) honeycomb \ce{MoS2} nanomeshes with the initial radius of circular holes: (b) R = 450 nm (H1). (c) R = 250 nm (H2). (d) R = 100 nm (H3). The calculated sensing parameters: (e) Summarized sensing response (\%); (f) The response time (s) from each device calculated at 2.5 ppb \ce{NO2} concentration; (g) The recovery time (s) from each device calculated from 2.5 ppb; (h) Sensitivity; (i) Limit of Detection (LOD).}
    \label{fig:3}
\end{figure}

The \ce{NO2} sensing response at 200$^\circ$C without humidity is shown in Figure~\ref{fig:3}. Here, the response ($R$, measured in \%) is defined as the ratio of change in the resistance upon exposure to \ce{NO2} ($\text{R}_{gas}$) to the base resistance in synthetic air ($\text{R}_{air}$): $R(\%) = (\frac{\text{R$_{gas}$-R$_{air}$}}{\text{R$_{air}$}}) \times 100$. The \ce{NO2} gas was injected for 600 s and then turned off to allow for recovery in the background environment for all measurements. The left axis of each graph displays the calculated variation in response, while the right axis shows the corresponding variation in resistance.
  
In Figure~\ref{fig:3}a-d, the baseline resistance is increased from 0.12 M$\Omega$ for the unpatterned device to 0.8 M$\Omega$ for H3. This is attributed to the concurrent rise in the $p$-doping effect due to the increased density of hexagonal holes in the honeycomb \ce{MoS2} nanomeshes, as discussed in SI. The observed increase in resistance with \ce{NO2} exposure suggests that \ce{NO2} is acting as an electron acceptor from $n$-type \ce{MoS2}. Moreover, the change in resistance due to \ce{NO2} exposure is also concurrently increasing with the density of honeycomb \ce{MoS2} nanomeshes. The obtained response to 2.5 ppb \ce{NO2} is 48\%, 64\%, 100\% and 120\% for devices H0, H1, H2, and H3, respectively. These results demonstrate that our devices are highly sensitive even to trace amounts of \ce{NO2}. The response value indicates fewer available \ce{NO2} adsorption sites in the unpatterned device, whereas in the nanomeshes, these sites' density increases, as shown in Figure~\ref{fig:3}e. When the devices are exposed to \ce{NO2}, the \ce{NO2} gas is accepting electrons from the $n$-type \ce{MoS2} surface due to its strong oxidative nature~\cite{cho2015highly, agrawal2021strategy}. As the hexagonal hole density increases from device H1 to H3, the number of zigzag edge sites also increases. The zigzag edges have higher chemical activity and strong gas adsorption ability in comparison to \ce{MoS2} basal plane (as we demonstrate in DFT section below). This makes them highly attractive sites for \ce{NO2} adsorption~\cite{jaramillo2007identification, yue2013adsorption}. Another interesting outcome observed in Figure~\ref{fig:3}a-d is the increase in resistance noise in devices H0 to H3. The low signal-to-noise ratio poses an unavoidable obstacle in the trace detection of gases. Taking this into consideration, we have selected sensor device H2 as the optimal nanopatterned device, combining high sensitivity with low noise, and unpatterned device H0 as the control device for future optimization. Sensing parameters, including response (\%), response time, and recovery time, are depicted in Figure~\ref{fig:3}e-g. The response time is determined by measuring the time interval between the sensor's transition from 10\% to 90\% of its response range. Similarly, the recovery time is characterized by the duration between the sensor's response shifting from 90\% to 10\% of its range. The response time ranges from 380 to 402 s, while recovery time falls in the range of 1500 to 3600 s for a 2.5 ppb \ce{NO2} cycle of each device. It is worth noting that the baseline resistance is nearly recovered to its initial level in synthetic air for each device at high operating temperatures, confirming the good recovery kinetics of each device. The sensitivity ($S$, measured in \%/ppb) is also an important factor and it is given by the slope of response (\%) and concentration ($C_{NO_2}$, ppb), $S = \frac{R(\%)}{C_{NO_2}}$.~\cite{david2020highly}
 
The calculated sensitivity profile is shown in Figure~\ref{fig:3}h. It is evident that for the unpatterned device H0, the sensitivity is the lowest while reaching the highest values for the honeycomb \ce{MoS2} nanomesh device H2, and saturating for device H3. The elevated sensitivity values imply a maximum change in response (\%) can occur per ppb \ce{NO2} exposure.  Moving on, we calculated the limit of detection (LOD) for each device, as illustrated in Figure \ref{fig:3}i. The LOD was calculated using the following equation: $\text{LOD} = \frac{3 \text{RMS$_{noise}$}}{S}$, where $S$ is the slope given by the response \textit{vs.} concentration curve. The RMS$_{noise}$ can be calculated using: $\text{RMS}_{noise} = \sqrt{\frac{\sum^{N}_{i=1}{(\text{R}_i - \bar{\text{R}})^2}}{N}}$. Additional details about fitting LOD data are provided in Figures~S3-S4. 

Remarkably, at 200$^\circ$C, the unpatterned device H0 exhibited the lowest LOD of 4 ppt, although its sensitivity was the lowest. At the same time, a honeycomb nanomesh device H2 showed an impressive LOD of 400 ppt combined with exceptional sensitivity, a notable achievement compared to the state-of-the-art (comparison between various sensing platforms is discussed below). The low LOD of H0 can be attributed to the high signal-to-noise ratio in the unpatterned device compared to honeycomb patterned \ce{MoS2} devices H1 -- H3.

\subsubsection{Humidity-tolerant and selective \ce{NO2} sensing at 200$^\circ$C}

\begin{figure}[ht!]
    \centering
    \includegraphics[width=0.99\linewidth]
    {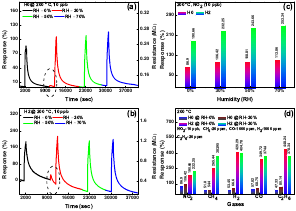}
    \caption{The dynamic \ce{NO2} sensing for 10 ppb at 200$^\circ$C for variable humidity ranging from 0\% to 70\% (a) Unpatterned device (H0). (b) Honeycomb \ce{MoS2} nanomesh device (H2). (c) The comparative sensing response (\%) for 10 ppb \ce{NO2} for unpatterned device (H0) and honeycomb nanomesh device (H2). The selectivity profile for the unpatterned device and honeycomb nanomesh device (H2) under 0\% and 30\% humidity.}
    \label{fig:4}
\end{figure}
 
Humidity is a critical issue in developing gas sensor applications. Typically, the baseline resistance is greatly affected by the presence and variation in humidity levels. Thus, real-world gas sensors are difficult to obtain under variable humidity~\cite{pei2021ti3c2tx, borini2013ultrafast, potyrailo2011materials}. Considering this, we have tested our sensors for the different RH levels, such as 0\%, 30\%, 50\%, and 70\%. The gas sensing performance for the unpatterned device H0 and honeycomb \ce{MoS2} nanomesh device H2 at 10 ppb \ce{NO2} and 200$^\circ$C against variable humidity has been assessed. As depicted in Figure~\ref{fig:4}a-b, both devices remain almost unaffected by variations in humidity levels. Interestingly, we observed that both devices exhibited slightly enhanced \ce{NO2} responses in the presence of humidity. When the 30\% humidity was inserted in the chamber, there was a sudden rise in the response, shown as the dotted circle for both devices in Figure~\ref{fig:4}a-b. However, with a further increase in RH levels, we did not observe any substantial variations in baseline resistance. The unpatterned device H0 maintained a stable baseline, while the honeycomb nanomesh device H2 exhibited a gradual decrease in baseline resistance over the same period due to faster recovery under humidity (for additional details, see SI, Section 6.0.6).


A bar graph comparison of calculated response (\%) for both devices is shown in Figure~\ref{fig:4}c. The honeycomb nanomesh device exhibited higher \ce{NO2} sensing performance under different humidity variations. The detailed sensing mechanism of \ce{NO2} interaction with unpatterned and honeycomb \ce{MoS2} nanomeshes under a humid environment is discussed in detail later in the mechanism section in SI.

Selectivity is another important aspect of developing a versatile gas sensor. In Figure~\ref{fig:4}d, we present the selectivity experiments for different gases such as \ce{NO2}, CH$_{4}$, H$_{2}$, CO, C$_{2}$H$_{6}$ under the 0\% and 30\% humidity. The unpatterned device H0 showed nearly the same response for all mentioned gases. However, the nanopatterned device H2 showed exceptional selectivity for \ce{NO2} under 0\% and 30\% humidity. Remarkably, the concentrations of other gases were in the 20 -- 1000 ppm range, while \ce{NO2} was present at the 10 ppb level. The difference between the responses therefore exceeds 1000 times in favor of \ce{NO2}. This confirms the selective nature of the honeycomb nanomesh devices for \ce{NO2} gas sensing. Based on these results, we conclude that our devices are highly sensitive and selective towards trace detection of \ce{NO2} even at high relative humidity levels (see Figure S5 for complete sensing cycles).

\begin{figure}[ht!]
    \centering
    \includegraphics[width=0.80\linewidth]
    {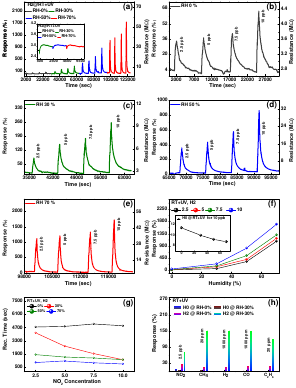}
    \caption{(a) The broad dynamic \ce{NO2} sensing response (2.5 -- 10 ppb) at RT under UV illumination obtained from the honeycomb nanomesh device (H2). Inset showed the change in resistance only due to variation in Humidity. The zoomed \ce{NO2} sensing  response for different humidity conditions (b) 0\% (c) 30\% (d) 50\% (e) 70\%. (f) The summarized sensing response (\%) for the honeycomb nanomesh device H2. The inset shows the sensing response (\%) for unpatterned device H0. (g) The calculated recovery time for each concentration with variable humidity. The selectivity profile of nanomesh and unpatterned \ce{MoS2} under 0\% and 30\% RH for various gases under RT with UV illumination.}
    \label{fig:5}
\end{figure}

\subsubsection{Room temperature \ce{NO2} sensing under UV illumination: excellent recovery and humidity-enhanced response}

\begin{figure}[ht!]
    \centering
    \includegraphics[width=0.99\linewidth]{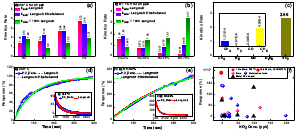}
    \caption{Calculated adsorption rate ($k_{\text{ads}}$) and desorption rate ($k_{\text{des}}$) using the Langmuir and Langmuir-Hinshelwood model for 2.5 ppb \ce{NO2} cycle. The desorption rate ($k_{\text{des}}$) is multiplied by a factor of 1000 for clarity. (a) At T = 200$^\circ$C. (b) At RT with humidity variation under UV illumination. (c) Fitted adsorption rate ($k_{\text{ads}}$) for different gases. The inset shows the magnified view of four different gases. (d-e) Experimental data fitting for the adsorption profile using Langmuir and L-H model for 2.5 ppb \ce{NO2} cycle for: (d) at T = 200$^\circ$C (inset shows the experimental data fitting for the desorption profile), and (e) at RT with 50\% humidity under UV illumination. (f) Performance comparison of this work with other relevant publications. The emphasis is given to reports where ppb levels of \ce{NO2} detection has been tested experimentally (see Table S1 for further details)}
    \label{fig:6}
\end{figure}

  
An alternative to the heating approach to counteract the effects of moderate humidity levels is the material surface activation by UV light. UV activation is efficient in forcing desorption of O$_{2}$ and H$_{2}$O from \ce{MoS2}, which facilitates interactions of \ce{NO2} with the material surface due to higher availability of adsorption sites~\cite{chen2001molecular, zhang2020optoelectronic}. We performed \ce{NO2} sensing experiments under continuous humidity exposure for a wide range of RH ranging from 0\% to 70\% (\ce{NO2} sensing results at RT without UV illumination, i.e. in the dark, is discussed in detail in SI and Figure S6). The broad sensing profile is shown in Figure~\ref{fig:5}a. As previously, we used four \ce{NO2} concentrations -- 2.5, 5, 7.5, and 10 ppb, tested at each humidity level. The grey, green, blue, and red curves denote 0\%, 30\%, 50\%, and 70\% humidity, respectively. The inset in Figure~\ref{fig:5}a shows the change in \ce{MoS2} resistance under the UV light illumination at different humidity levels. It can be seen that the baseline resistance initially drops but quickly recovers (on the period of $\sim$20 min). Thus, the effect of humidity is mitigated by the UV illumination with time. The baseline remains stable even under harsh rises in RH, confirming the potential use of \ce{MoS2} nanomeshes for real-world gas sensing applications. Furthermore, Figure~\ref{fig:5}b-e illustrates a humidity-enhanced \ce{NO2} sensing. Specifically, the response to 2.5 ppb \ce{NO2} is increased 33-fold (from 34\% to 1120\%) upon RH rise from 0\% to 70\%. The achieved response values were extremely high for such a low \ce{NO2} concentration with humidity variation, see Figure~\ref{fig:5}f. They outperform the highest responses at such low \ce{NO2} concentration previously reported for any \ce{MoS2}-based sensors (see Figure~\ref{fig:6}f and Table S1 for detailed comparison). Additionally, the recovery time decreased nearly 6-fold, from 5000 s to 800 s, upon increasing the humidity level, as shown in Figure~\ref{fig:5}g. We have also tested the unpatterned device H0. These measurements are shown in Figure S7. The inset of Figure~\ref{fig:5}f shows the \ce{NO2} response under the identical set of humidity conditions for unpatterned device H0, which showed a decreased response with humidity and an overall poor performance. Since the baseline resistance fluctuation is very high due to variation in humidity in this case, we consider the relative response (\%) of the signal. The relative response is determined by subtracting the response value when \ce{NO2} is turned on from the response value achieved when \ce{NO2} is off. The baseline resistance was not recovered under UV exposure of unpatterned device H0, unlike the nanomesh device H2, which showed both complete and speed-up recovery. Furthermore, we tested selectivity for various gases under the humid environment (see Figure~\ref{fig:5}h). We observe improved gas sensing performance for all the studied gases. However, the patterned device H2 exhibits an especially selective and sensitive behavior upon exposure to \ce{NO2}. These results further confirm the room temperature, UV-activated, and humidity-enhanced performance of our chemiresistive gas sensor, which paves the way towards real-world gas sensing applications under high humidity conditions. 

To understand the role of zigzag edges in honeycomb nanomeshes for \ce{NO2} adsorption, we studied the adsorption and desorption kinetics using the Langmuir and Langmuir-Hinshelwood (L-H) models. We start with discussing the Langmuir model. The adsorption and desorption rate kinetics were calculated for 2.5 ppb \ce{NO2} cycle of each device (H0, H1, H2, and H3) at 200$^\circ$C and for device H2 only with humidity variation from 0\% to 70\% using the following two equations~\cite{koo2019catalytic}. The desorption kinetics is expressed as:

\begin{equation}
R(t) = R_0 \exp\left(-k_{\text{des}} t\right),
\label{Eq:des_kin}
\end{equation}

while the adsorption kinetics is:

\begin{equation}
R(t) = R_{\text{max}} \frac{C_{a}K}{1 + C_{a}K} \left(1 - \exp\left[-\left(\frac{1 + C_{a}K}{K}\right)k_{\text{ads}} t\right]\right).
\label{Eq:ads_kin}
\end{equation}

In the above equations, $R_{\text{max}}$ is the maximum response of the sensor for the cycle, $R_{0}$ is the response of the sensor during the recovery cycle when \ce{NO2} gas is turned off. $k_{\text{ads}}$ and $k_{\text{des}}$ are the adsorption and desorption kinetic rates, respectively. $K$ is the adsorption capacity of the sensor, defined as the ratio of $k_{\text{ads}}$ and $k_{\text{des}}$, $K = k_{\text{ads}}/k_{\text{des}}$. $C_a$ is the applied \ce{NO2} concentration (2.5 ppb in our calculations). The calculated parameters are depicted in Figure~\ref{fig:6}a-b. At 200$^\circ$C, the $k_{\text{ads}}$ and $k_{\text{des}}$ increase with an increase in the density of hexagons for patterned devices. The higher $k_{\text{ads}}$ means stronger interaction of \ce{NO2} gas molecules with the sensing surface. Notably, $k_{\text{ads}}$ is 1.86 times higher for honeycomb hexagon device H2 (2.66 ppm$^{-1}$s$^{-1}$) than the unpatterned device H0 (1.45 ppm$^{-1}$s$^{-1}$). The enhanced $k_{\text{ads}}$ reveals the importance of edges in superior \ce{NO2} gas sensing. Initially, $k_{\text{des}}$ is high for the unpatterned device H0; however, it decreases for the honeycomb hexagon device H1, indicating strong adsorption of \ce{NO2} gas molecules in the presence of edges. The value of $k_{\text{des}}$ is further increased for devices H2 and H3 due to the higher availability of favorable sites for \ce{NO2} adsorption. The fitting has been performed assuming that \ce{NO2} follows the mass action law, with the response being proportional to the amount of adsorbed \ce{NO2} gas. The desorption energy and charge transfer values of \ce{NO2} on patterned devices may differ from those on unpatterned devices, as suggested by the density functional theory calculations presented below. This can also be inferred from the reduced $k_{\text{des}}$ rate when transitioning from device H0 to H1. The patterned devices have similar adsorption and desorption energies but offer a higher number of adsorption sites for \ce{NO2}. Consequently, the $k_{\text{des}}$ rate increases further from device H1 to H3.

In Figure~\ref{fig:6}a, $k_{\text{des}}$ for each device have been multiplied by a factor of 1000 for clarity. We note that the Langmuir model fits our experimental findings very well. Therefore, we extended the calculations to \ce{NO2} sensing under variation of humidity and with UV illuminations. However, we observed that $k_{\text{ads}}$ rate is decreased with an increase in humidity while $k_{\text{des}}$ rate is increased. The calculated $k_{\text{des}}$ rate for hexagonal honeycomb device H2 under 70\% humidity (3.47${\times}$10$^{-3}$ s$^{-1}$) is 4.11 times higher than the 0\% humidity (0.84${\times}$10$^{-3}$ s$^{-1}$). This confirms the improved recovery under humidity with UV illumination. However, at the same time, the $k_{\text{ads}}$ rate for hexagonal honeycomb device H2 under 70\% humidity (0.16 ppm$^{-1}$s$^{-1}$) is 9.68 times lower than the 0\% humidity (1.56 ppm$^{-1}$s$^{-1}$). The \ce{NO2} response is the highest under 70\% RH and with UV illumination which implies that the Langmuir model is not ideal to fit the experimental findings under humidity.

In the Langmuir model, the $k_{\text{ads}}$ and $k_{\text{des}}$ can be further explained in terms of activation energy at a particular temperature~\cite{zhang2013room, gu2020visible}. Furthermore, when the sensor is exposed to light, the adsorption and desorption constants start to depend on the illumination conditions. Additionally, humidity significantly affects the kinetic rates, which needs to be taken into account during the fitting. Hence, a complex interplay between these factors may explain why the Langmuir model does not satisfactorily fit the experimental data under humidity and UV illumination.

To gain further insight into the kinetic processes, we employed the L-H model for 200$^\circ$C and variable humidity data. The L-H model can be directly obtained from the Langmuir model provided the condition $C_aK>>1$ is satisfied (estimated values of $C_aK$ in our case range from 1.5 to 5, depending on the experimental conditions). Thus, the L-H model effectively decouples the desorption process from the adsorption kinetics. With this in mind, the adsorption profile is fitted almost identically by both the L-H and the Langmuir models for devices H0, H1, H2, and H3 at 200$^\circ$C and device H2 at RT with variation of humidity from 0\% to 70\% (see Figure \ref{fig:6}a,b). The adsorption profile has been fitted using the following equation~\cite{lee2005understanding}:

\begin{equation}
R(t) = R_{\text{max}} \left(1 - e^{-k_{\text{ads}} C_{a} t}\right),
\end{equation}

where, $R(t)$ is the time-dependent response. $R_{\text{max}}$ is the maximum response and $C_{a}$ is the NO$_2$ concentration (which is 2.5 ppb). The fitted $k_{\text{ads}}$ values are displayed on the bar graph in Figure~\ref{fig:6}a-b. The L-H model adequately fits the experimental data of hexagonal honeycomb device H2 under variable humidity with UV illumination and shows a slight increase in the adsorption rate with an increase in the hexagon density. This contrasts with the Langmuir model, which in this case showed the opposite trend (see Figure~\ref{fig:6}b). The experimental data fittings of the adsorption rates using the Langmuir model and L-H model at 200$^\circ$C and at RT under humidity variation with UV illumination are displayed in Figure~\ref{fig:6}d-e. The insets of Figure~\ref{fig:6}d-e additionally show the desorption rates fittings.

To test the specificity of our gas sensors, the adsorption kinetic rates for different gases, including \ce{CH4}, \ce{H2}, CO, \ce{C4H6}, and \ce{NO2} were obtained using the Langmuir model, see Figure~\ref{fig:6}c. The corresponding $k_{\text{ads}}$ for \ce{NO2} is 2.66 ppm$^{-1}$s$^{-1}$, which is at least three orders of magnitude higher than the other measured gases. This confirms the ultra-selective nature of hexagonal honeycomb device H2 for \ce{NO2} gas.

Furthermore, we compared our results with other reports where ppb level \ce{NO2} detection has been tested experimentally. This comparison is displayed in Figure~\ref{fig:6}f. The comparison has been carried out using certain sensor categories, including bare \ce{MoS2}, \ce{MoS2} heterostructure with other materials, and metal nanoparticle doped \ce{MoS2}. The corresponding data of response (\%), concentration, and temperature from each reference is shown in Table S1. We note that our honeycomb hexagonal sensors, which fall into the category of bare \ce{MoS2} structures, exhibit superior performance compared to other \ce{MoS2}-based sensors, even those with more complex configurations like metal nanoparticle doping and heterostructures.  

\subsubsection{Density function theory calculations}

\begin{figure}[ht!]
    \centering
    \includegraphics[width=0.99\linewidth]{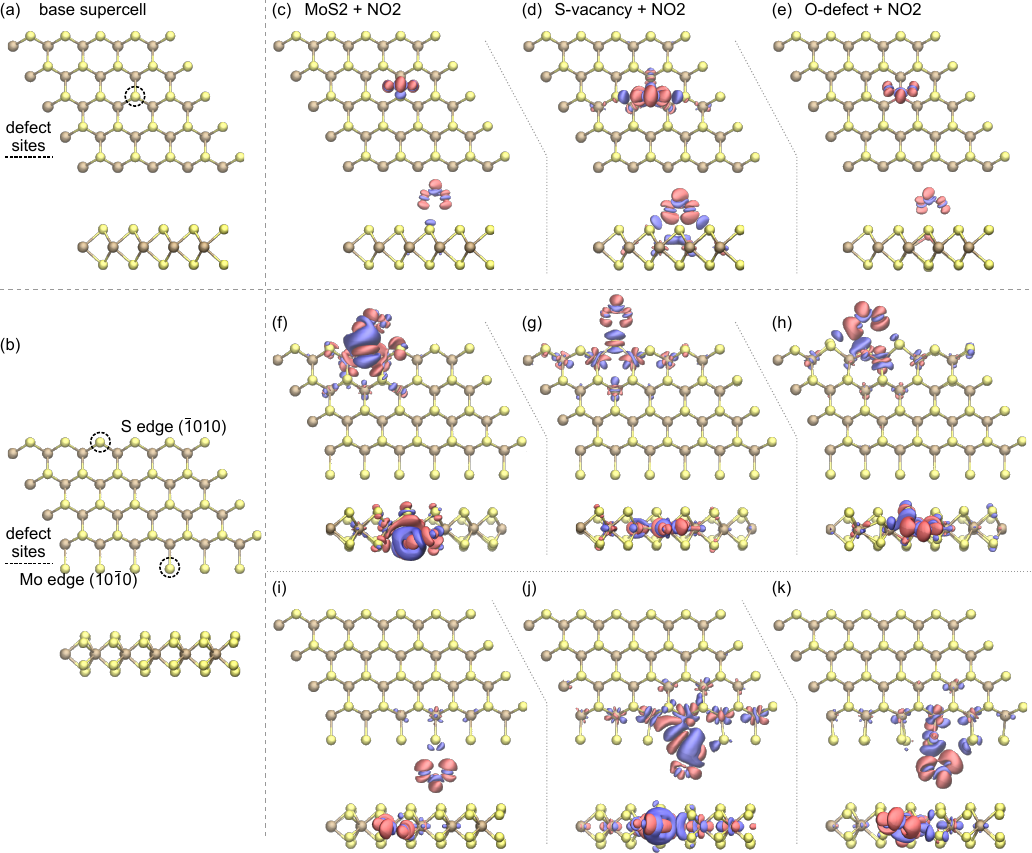}
    \caption{Geometry and electron charge density difference for the considered structures. The pristine \ce{MoS2} structures are sketched in (a) for the $5\times5$ supercell of \ce{MoS2} monolayer and (b) the $6\times4$ ribbon supercell. The dashed circles mark the defect placement, though only one defect is present at a time for the ribbon. (c-k) Isosurface plots of the electron charge density difference for \ce{NO2} adsorbed on \ce{MoS2} with top and side views. Charge accumulation is presented in red and depletion in blue.}
    \label{fig:dft}
\end{figure}

In this section, we complement the measurements with first-principle calculations (see Methods for details) based on density functional theory (DFT) of adsorption energies and charge transfer to an \ce{NO2} molecule adsorbed on \ce{MoS2}. These calculations are performed on both the basal plane supercell and the one-dimensional ribbon supercell, considering the three scenarios most relevant to our study. Specifically, we examine \ce{NO2} adsorption on pristine \ce{MoS2}, at a sulfur vacancy defect site (S-vacancy site), and at a sulfur vacancy site with an oxygen substitution (O-defect site). The latter two scenarios are particularly pertinent to our research, as the tested sensing devices are highly susceptible to sulfur vacancy generation due to the fabrication processes, which include both dry and wet chemical etching. Furthermore, these S-vacancy sites are highly prone to oxygen adsorption from the environment. The honeycomb \ce{MoS2} nanomesh can be considered as hexagonal nanoribbons; this is especially clear for thin \ce{MoS2} structures shown in Figure~\ref{fig:1}b-d. For computational efficiency, we consider only monolayers either in the form of a $5\times5$ supercell or a one-dimensional $6\times4$ ribbon supercell. The structures are shown in Figure~\ref{fig:dft}a,b respectively, where we marked the placement of the defect as the dotted circle. Thus, in total, we consider 9 adsorption geometries. First, these include \ce{NO2} adsorption above the basal plane of a pristine structure, above an S-vacancy and an O-defect. Second, for the \ce{MoS2} ribbon, we consider these three cases next to either of the zigzag edges: the Mo edge ($10\bar{1}0$) terminated with S dimers or a S edge ($\bar{1}010$)~\cite{PhysRevLett.87.196803}. An \ce{NO2} molecule was placed next to the marked site and the structure was relaxed using the optPBE-vdW functional until the maximum force on each atom was less than \unit[0.02]{eV/\AA}.

\begin{table}[H]
\centering
\caption{Comparison of binding energies $\Delta E$ and charge transfer $\Delta q$ to the adsorbed \ce{NO2} molecule at a particular site on \ce{MoS2}: ideal -- pristine \ce{MoS2} sheet; S-vac -- sulphur vacancy; O-def -- oxygen defect; for (top) a $5\times5$ supercell and (bottom) $6\times4$ zigzag ribbon with a Mo edge ($10\bar{1}0$) terminated with S dimers and a S edge ($\bar{1}010$).\cite{PhysRevLett.87.196803}} 
\label{tbl-1}
\begin{tabular}{ C{2.0cm} C{1.0cm} C{1.0cm} C{1.0cm} C{0.2cm} C{1.0cm} C{1.0cm} C{1.0cm} }
\toprule
 & \multicolumn{7}{l}{$5\times5$ monolayer supercell} \\
 & ideal & S-vac & O-def &  &  &  & \\
\cline{2-4}
$\Delta E$ (meV) & -263  & -235  & -287  & & & & \\
$\Delta q$ ($e$) & 0.12 & 0.26 & 0.11 & & & & \\
\toprule
 & \multicolumn{7}{l}{$6\times4$ monolayer ribbon}  \\
 & \multicolumn{3}{c}{Mo edge ($10\bar{1}0$)} &  & \multicolumn{3}{c}{S edge ($\bar{1}010)$} \\
 & ideal & S-vac & O-def &  & ideal & S-vac & O-def \\
\cline{2-4} \cline{6-8}
$\Delta E$ (meV) & -243  & -260  & -357  & & -1050 & -196  & -382  \\
$\Delta q$ ($e$) & 0.21 & 0.57 & 0.31 & & 0.53 & 0.12 & 0.26 \\
\bottomrule
\end{tabular}
\end{table}

The final geometries with relaxed \ce{NO2} adsorption are plotted in Figure S8. We focus on two important parameters: adsorption energy and charge transfer. The adsorption energy of each case is calculated as $\Delta E = E_\mathrm{complex} - E_\mathrm{MoS_{2}} - E_\mathrm{NO_{2}}$, with negative values indicating adsorption is energetically favorable. Here the obtained adsorption energies are static, i.e. at \unit[0]{K} with the zero point and thermal energies not included, as these have been shown to be small in comparison to the vdW contributions~\cite{rana2012comparing}. The energies are reported in Table~\ref{tbl-1} and all are significantly negative~\cite{CPL595-35}, indicating efficient adsorption of \ce{NO2} in all cases. We further analyze the charge redistribution in the systems upon \ce{NO2} adsorption based on the Bader charges \cite{BaderAnalysis}. The positive value indicates that the charge is transferred from \ce{MoS2} to \ce{NO2}. The results confirm that \ce{NO2} plays the role of an acceptor, in every case taking up a sizable amount of charge from \ce{MoS2}, as illustrated in Table~\ref{tbl-1}. This charge transfer is also visualized in Figure~\ref{fig:dft}, where charge accumulation can be seen at the position of the \ce{NO2} molecule with an accompanying depletion in the \ce{MoS2}. For the ribbon configuration, we analyze both Mo- and S-terminated edges, which are predominantly located at the perimeter of the honeycomb pattern. Consequently, S-vacancy and O-defect sites emerge as the most relevant at the hexagonal perimeter. Our devices were fabricated from a 2H-\ce{MoS2} crystal having AB stacking, which suggests the presence of both types of defect sites in multilayer \ce{MoS2} honeycomb nanomesh structures. This stacking configuration further enhances the potential for diverse defect interactions, contributing to the overall sensing performance. Achieving ideal pristine Mo and S edges and defect-free basal plane supercells is nearly impossible in practical settings due to the inherent limitations of our fabrication processes, which involve both dry and wet etching techniques. Thus, it is likely that in real experiments, the most favorable adsorption sites are S-vacancies and O-defects on one-dimensional edges. The calculated charge transfer is nearly doubled for one-dimensional edges for both Mo- and S-terminated edges in comparison to the basal plane. Moreover, the binding energy is higher for O-defects in all cases.

We now turn our attention to the correlation between the DFT results and our experimental findings. For chemiresistive sensors, charge modulation due to exposed gas molecules is a crucial factor, as it ultimately translates into an electrical signal. A higher charge transfer from \ce{MoS2} to \ce{NO2} results in an enhanced \ce{NO2} sensing response. At room temperature, we find that O-defect sites are predominantly available in the basal plane supercell and in both Mo- and S-terminated ribbon configurations. The combined charge transfer of both edges is substantially higher for the honeycomb \ce{MoS2} nanomesh structure compared to the O-defect sites of the basal plane. Our RT \ce{NO2} sensing experiments conducted without UV activation corroborate this enhanced performance. Specifically, the honeycomb \ce{MoS2} nanomeshes (H2) exhibited nearly doubled \ce{NO2} sensing response compared to the unpatterned device (H0), albeit with incomplete recovery. Furthermore, under thermal or UV activation, weakly adsorbed oxygen molecules are desorbed, facilitating \ce{NO2} adsorption at the most favorable sites. Specifically, the S-vacancy and O-defect sites at the edges of honeycomb \ce{MoS2} nanomesh devices exhibit charge transfer values of 0.57$e$ and 0.31$e$ (at Mo-edge), respectively, making them highly effective for \ce{NO2} sensing.

\subsubsection{Discussion of sensing mechanism}

In this study, we employed \ce{MoS2} as the active sensing material for the detection of \ce{NO2}. \ce{MoS2} nanomeshes were fabricated using flakes mechanically exfoliated from a commercial $n$-type 2H-\ce{MoS2} crystal~\cite{HQ}. Due to dry and wet etching, the resulting nanomeshes acquire a certain level of $p$-doping, which does not change the overall $n$-type nature of \ce{MoS2}. The operational principle underlying the detection of \ce{NO2} involves the perturbation of charge carriers (hence, resistance) in \ce{MoS2} due to the surface charge transfer interactions between the \ce{MoS2} and \ce{NO2} molecules. The \ce{NO2} concentration, flow rate, and the \ce{MoS2} surface stoichiometry influence the sensor's resistance. The interaction between \ce{NO2} and \ce{MoS2} varies under different operating conditions. Therefore, the sensing mechanism involves a complex interplay of physical and chemical processes, making it challenging to define precisely. Given this complexity, a comprehensive understanding of the sensing mechanism requires further investigation beyond the scope of this study. In this section, we briefly summarize the critical aspects of \ce{MoS2} nanomeshes sensing behavior and suggest directions for future research. Detailed discussion is provided in the SI (section "Sensing mechanism").

The oxidative wet etching and environmental factors (gas atmosphere, humidity, temperature, light) significantly alter the \ce{MoS2} surface, potentially capping edges and sulfur vacancies with oxygen molecules or even forming O-defect sites. The nanomesh \ce{MoS2} devices (H1, H2, and H3) feature an abundance of edge sites, possibly capped with oxygen. Enhanced \ce{NO2} adsorption and charge transfer at the \ce{MoS2} edges, particularly in the honeycomb H2, improve device response, affecting resistance even at single-digit ppb \ce{NO2} levels. 

In dry air and in the dark, molecular oxygen likely dominates the surfaces of all proposed devices, thereby diminishing the adsorption sites for \ce{NO2}. Despite this, the 2.5 ppb \ce{NO2} response in the patterned device H2 reaches 8\%, exceeding the 4\% value for the unpatterned device H0. Additionally, the strong binding between \ce{NO2} and \ce{MoS2} leads to incomplete recovery at room temperature in the dark, as illustrated in Figure S9.

To achieve efficient and reversible \ce{NO2} detection, we first raise the temperature to 200$^\circ$C. \ce{NO2} has higher binding energy to \ce{MoS2} than \ce{O2}, so the elevated temperature facilitates the removal of oxygen adsorbed on the surface and creates more available sites for \ce{NO2} adsorption. At 200$^\circ$C, \ce{MoS2} nanomeshes show a significantly higher response than the unpatterned device H0, indicating that the edge sites are advantageous for \ce{NO2} sensing. Moreover, the elevated temperature allows for full sensor recovery, which is not possible at room temperature.

In dry air, our second approach leverages UV illumination to both remove the adsorbed oxygen from the surface and enhance \ce{NO2} adsorption. The photogenerated electron-hole pairs are crucial in modulating the adsorption and desorption thermodynamics and kinetics of both O$_2$ and \ce{NO2} molecules. This effect is especially pronounced in the \ce{MoS2} nanomesh H2, which exhibits higher \ce{NO2} sensitivity than its H0 counterpart. Therefore, UV illumination is critical for amplifying \ce{NO2} detection. 

In dark and highly humid environments, the \ce{NO2} sensing mechanism becomes complex, mainly due to the significant influence of water vapor, as elaborated in the SI. The presence of water vapor alters \ce{NO2} detection through interactions with surface-bound hydroxyl ions (OH$^{-}$) and protons (H$^{+}$). These interactions may lead to the formation of surface-bound nitrate ions (NO$_3$$^{-}$(ads)), which are characterized by their elevated binding energy and enhanced charge transfer on \ce{MoS2}~\cite{yao2011humidity, han2019interface}. This process enhances the ability to detect \ce{NO2} in humidity-rich conditions, even in the dark. 

The most compelling results of our investigation were achieved under high humidity and UV illumination at RT. Under these conditions, the \ce{MoS2} nanomesh H2 device demonstrated a universal enhanced sensing response across all tested gases, including \ce{NO2}, CH$_4$, H$_2$, CO, and C$_2$H$_6$. In contrast, the unpatterned device H0 showed degraded performance when exposed to humidity, highlighting the exceptional role of zigzag edges in the H2 device's enhanced gas sensing capabilities. This universal behavior suggests that the adsorption and charge transfer of all gases is higher in the presence of humidity for \ce{MoS2} nanomesh device H2. Furthermore, it implies that the nanomesh surface has the most available sites for gas molecule adsorption.

While the precise mechanism behind the enhanced sensing performance remains elusive, we hypothesize that the synergy between humidity and UV light optimally cleans the surface. Specifically, UV illumination plays a dual role, removing adsorbed oxygen and preventing the accumulation of surface-bound hydroxyl ions (OH$^{-}$) and protons (H$^{+}$). This hypothesis is supported by the observed stable baseline resistance even at high RH levels at RT. Therefore, the exceptionally clean surface of our \ce{MoS2} nanomesh H2 device significantly enhances its sensing response across all tested gases. Additionally, we can not rule out the possibility of enhanced adsorption and charge transfer between the exposed gas molecules and \ce{MoS2} nanomesh H2 surface in the presence of adsorbed water. The enhanced \ce{NO2} adsorption and charge transfer in the presence of humidity has been previously reported using DFT~\cite{azizi2020high}.  
\subsection{Conclusion}
In conclusion, we have successfully demonstrated the fabrication of controlled, uniform, and sharp zigzag edge enriched hexagons \ce{MoS2}. We developed room-temperature ultra-sensitive, ultra-selective, and humidity-enhanced \ce{NO2} gas sensors. The sensor response to 2.5 ppb \ce{NO2} reaches 99\% at T=200$^\circ$C and 1120\% at room temperature under UV illumination with 70\% relative humidity. We showed that honeycomb \ce{MoS2} exhibits superior \ce{NO2} sensing capability due to the high number of zigzag edges, as supported by DFT calculations. Our devices achieved an impressive parts-per-trillion detection limit even at high relative humidity. Our work lays the foundation for developing nanostructured \ce{MoS2} as real-life \ce{NO2} sensors operable at room temperature, under UV light illumination, and in harsh humidity conditions.

\subsection{METHODS AND MATERIALS}
\subsubsection{Electrical Contacts Fabrication}
The \ce{MoS2} flakes were mechanically exfoliated from HQ-graphene crystals using polydimethylsiloxane (PDMS) stamps and then transferred onto $n$-doped  285 nm SiO$_{2}$/Si substrate. The different hexagon devices were fabricated onto identical thickness flakes. The thickness of the flakes was measured using the VEECO profilometer. Finally, the Autocad software was used to design the circular holes and contacts. The flakes with substrates were first covered with a thick layer of 300 nm of 950 PMMA A4 photoresist (MicroChem, USA) using the spin coating technique. The e-beam lithography was employed using the JEOL JBX 9300FS lithographer at 100 kV accelerating voltage and 30 nA current. The lithographically patterned design was developed using 1:3 MIBK:IPA mixture for 2 min and 20 s. The Cr/Au (5/200 nm)metals were evaporated by Lesker PVD 225 e-beam evaporation system (Kurt J. Lesker Company, Germany), followed by overnight liftoff in acetone.
 
\subsubsection{Honeycomb Mesh Nanopatterning}
After the device fabrication, we covered the device with a 500 nm thick layer of ARP 6200.13 e-beam resist (Allresist GmbH, Germany). Again, the e-beam lithography was employed using the JEOL JBX 9300FS at 100 kV accelerating voltage and 2 nA current. The lithographically patterned design was developed using $n$-amyl acetate for 1 minute 15 s. The reactive ion etching has been performed in Oxford Plasmalab 100 system (U.K.) using CHF$_{3}$ plasma. The dry etching conditions were as follows: CHF$_{3}$ and carrier gas (argon) flow were 50 sccm and 40 sccm, respectively. The radio frequency power was 50 W. The \ce{MoS2} etching rate was $\sim$ 10 nm/min. The devices were cleaned using 1165 remover, acetone, and isopropyl alcohol (IPA) for 5 min each. The wet etching process was performed in the mixture of  H$_2$O$_2$:NH$_4$OH:H$_2$O with a volumetric ratio of 1:1:10 at a slightly elevated temperature~\cite{munkhbat2020transition}.

\subsubsection{Instrumental Characterization}
The SEM imaging was performed using the Ultra 55 FEG SEM at the acceleration voltage of 3 and 4 kV. The Raman spectra were collected at room temperature using the WITEC Alpha 300R equipped with a 532 nm laser with 1800 l/mm grating. 

\subsubsection{Gas Sensing Experiments}
The gas sensing measurements were performed in a home-built sealed chamber, using 20\% O$_{2}$ and 80\% N$_{2}$ (dry synthetic air) as the background gas. The source bottle, which was used to produce \ce{NO2} pulses, contained 100 ppb \ce{NO2} in a background of \ce{N2}, with a purity of N6.0. Gas mixing was performed using a gas mixer setup comprising low-flow mass flow controllers (Bronkhorst). The lower \ce{NO2} concentrations were achieved by diluting the original 100 ppb \ce{NO2} using pure N$_{2}$. The overall volume of the sensing environment was always fixed at 100 ml/min. The \ce{NO2} gas was on during measurement for 600 s. The 355 nm UV LED was used for testing the device under UV light. The sensing temperature was controlled using a ceramic heater (Heraues GmbH, Germany). The electrical testing has been done using the single channel source measure unit (SMU, Keithley 2601 SourceMeter).

\subsubsection{DFT calculations}
The DFT calculations we carried out using the open-source GPAW package \cite{2005_PRB_71_035109_mortensen, 2010_JPhysCondMat_22_253202_enkovaara} which utilizes the projector-augmented wave method. We used the optPBE-vdW functional to treat the exchange-correlation potential and van der Waals interactions \cite{libvdwxc}. For the basal plane calculations, we used a $5\times5\times1$ supercell of 1H-\ce{MoS2} with \unit[15]{\AA} of vacuum above and below the monolayer to avoid interlayer interactions. 
The Kohn-Sham orbitals were expanded using a plane wave basis set with an energy cutoff of 500~eV, we used a $5\times5\times1$ Monkhorst-Pack grid for the $\mathbf{k}$-mesh and a grid spacing of less than 0.2~\AA.
The \ce{MoS2} ribbon was constructed from a pristine $6\times5$ monolayer by removing the first row of Mo atoms, leaving an S-terminated ribbon on both sides. The vacuum for the ribbon was set to \unit[10]{\AA} above and below, as well as to either side of the edges. For these calculations, we used a $5\times1\times1$ Monkhorst-Pack grid for the $\mathbf{k}$-mesh, while the other parameters were the same.
For all calculations, we used  Fermi-Dirac occupation number smearing with a factor of 20~meV and line-shape broadening of 50~meV.

Initially, we relaxed the pristine $5\times5$ monolayer and the $6\times4$ ribbon until the maximum force on each atom was smaller than \unit[5]{meV/\AA}. Next, the relaxed structures were modified by introducing the two types of defects. These structures were subsequently relaxed using the same conditions. The final relaxed structures (with and without defects) had \ce{NO2} added to the vicinity of the defect (or, respectively, where it would be) and relaxed until the forces were less than \unit[0.02]{eV/\AA}.


\begin{suppinfo}
Detailed experimental procedures, supplementary notes, and
additional characterization data.
\end{suppinfo}
\begin{acknowledgement}
This work was performed in part at Myfab Chalmers and Chalmers Materials Analysis Laboratory (CMAL). A.V.A. is thankful to SIO Grafen’s 2D Graduate network for the mobility grant to conduct experiments at Linköping University. A.V.A, A.Yu.P, and T.O.S. acknowledge funding from the Olle Engkvist Foundation (Grant No. 211-0063), 2D-TECH VINNOVA competence center (Ref. 2019-00068), Chalmers Excellence Initiative Nano, and Knut and Alice Wallenberg Foundation (KAW, grant No. 2019.0140). T.J.A. acknowledges funding from the Polish National Science Center, project 2019/34/E/ST3/00359. 
\end{acknowledgement}

\section*{Conflicts of interest} The authors declare no conflicts of interest.

\section*{Keywords} nitrogen dioxide (\ce{NO2}) gas sensing, nanostructured molybdenum disulfide (\ce{MoS2}), transition metal dichalcogenides, relative humidity, chemiresistive gas sensing

\newpage
\bibliography{ManuscriptGasSensing}

\newpage
\subsubsection{TOC Graphic}
\begin{figure}[ht!]
    \centering
    \includegraphics[width=0.99\linewidth]
    {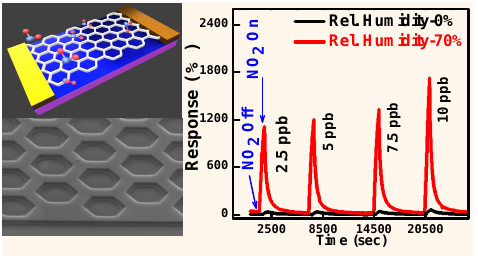}
    \label{fig:TOC}
\end{figure}

\end{document}


\maketitle


\newpage
\tableofcontents

\newpage

\section{Supplementary Figures}

\begin{figure}[ht!] 
    \centering
    \includegraphics[width=0.99\linewidth]
    {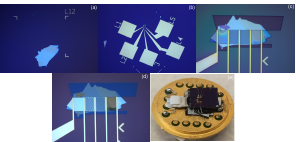}
    \caption{Fabrication process from exfoliation to final device fabrication. Optical images of: (a) Mechanically exfoliated flakes, (b) Cr/Au contact deposition, (c) Dry etching process on the fabricated devices, (d) Final devices after wet etching process. (e) Photograph of devices mounted on top of the Pt100 heater.}
    \label{fig:S1 }
\end{figure}

\newpage
\section*{Electrical characteristics of MoS$_2$ nanomeshes in ambient air}
 
\begin{figure}[ht!]
    \centering
    \includegraphics[width=0.99\linewidth]
    {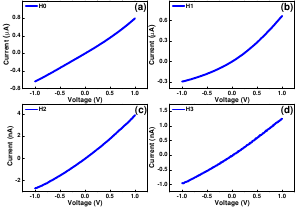}
    \caption{Current \textit{vs.} voltage for (a) unpatterned device (H0) and honeycomb MoS$_{2}$ nanomeshes of initial circular etching pits: (b) R-450 nm (H1),  (c) R-250 nm (H2), and (d) R-100 nm (H3) at room temperature.}
    \label{fig:S2}
\end{figure}

We assess the current \textit{vs} voltage (\textit{I-V}) characteristics for each device discussed in the main text. The \textit{I-V} characteristics in ambient air at room temperature (RT) are illustrated in Figure~\ref{fig:S2}. This characteristic distinctly reveals a reduction in current with an increase in hexagon density. Notably, the original MoS$_2$ crystal employed in this study exhibits $n$-type behavior. The decrease in current suggests the extraction of electrons from each device, attributed to the fabrication process and the heightened hexagon density. The highly dense honeycomb MoS$_{2}$ nanomesh device H3 exhibited the minimum current compared to the unpatterned device H0. Our device fabrication process involves combination of dry etching (CHF$_{3}$ plasma) and wet etching (H$_{2}$O$_{2}$:NH$_{4}$OH:H$_{2}$O) techniques.

Both of these techniques played significant roles in extracting electrons from the MoS$_{2}$. The CHF$_{3}$ plasma can introduce efficient $p$-doping~\cite{wi2014enhancement, chen2013stable, azcatl2014mos2}. CHF$_{3}$ is a fluorine (F)-containing plasma characterized by strong electronegativity, enabling the feasible extraction of electrons from MoS$_{2}$. Yue \textit{et al.} conducted theoretical calculations, indicating that F-dopants can extract 0.675$e$ from MoS$_{2}$~\cite{yue2013functionalization}. It is noteworthy that, in the context of the basal plane of MoS$_{2}$, the top layers are typically affected gradually by F-dopants. Wi \textit{et al.} performed the angle-resolved X-ray photoelectron spectroscopy (ARXPS) to study the doping depth by the CHF$_{3}$. The doping depth was found to be the highest for angles greater than 30$^{\circ}$ relative to the surface normal, reaching an effective depth of 6.5 nm. Notably, in our case, circular etching pits were completely drilled. Therefore, in our case, F-dopants can uniformly extract electrons even from the bottom layers. Consequently, we observed a significant increase in $p$-doping with an increase in the etching pits density. We have also observed a decrease in current confirming similar behavior. (The data is not shown) Next, we have done wet etching using the solution mixture of  1:1:10 H$_{2}$O$_{2}$:NH$_{3}$:H$_{2}$O. The combination of H$_{2}$O$_{2}$ and NH$_{4}$OH serves to further extract electrons, resulting in an increased generation of $p$-doping in the devices~\cite{gan2016photoluminescence, su2015tuning}. It is worth noting that the unpatterned device H0 was completely covered with the resist during dry etching. However, the unpatterned device H0 and all other devices underwent exposure to wet etching treatment. Another crucial point to note is that at the zigzag edges, the adsorption of environmental oxygen (O$_{2}$) is highly probable. The adsorbed O$_{2}$ can significantly extract electrons from the edges~\cite{yue2013adsorption, nan2014strong}.

\newpage
\section{Theoretical limit of detection (LOD) calculation}
\begin{figure}[ht!]
    \centering
    \includegraphics[width=0.99\linewidth]
    {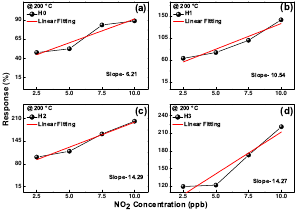}
    \caption{Linear fit of response (\%) \textit{vs.} concentration for each device, (a) unpatterned device (H0) and honeycomb MoS$_{2}$ nanomeshes of initial circular etching pits: (b) R--450 nm (H1),  (c) R--250 nm (H2), and (d) R--100 nm (H3).}
    \label{fig:S3}
\end{figure}

\begin{figure}[ht!]
    \centering
    \includegraphics[width=0.99\linewidth]
    {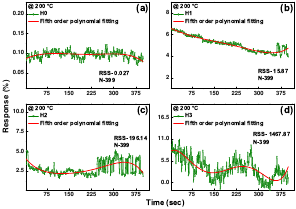}
    \caption{Fifth order polynomial fit of response (\%) for each device, (a) unpatterned device (H0) and honeycomb MoS$_{2}$ nanomeshes of initial circular etching pits: (b) R--450 nm (H1), (c) R--250 nm (H2), and (d) R--100 nm (H3).}
    \label{fig:S4}
\end{figure}

The theoretical limit of detection (LOD) has been calculated from experimental data as per the International Union of Pure and Applied Chemistry (IUPAC) definition and reported protocols
~\cite{wang2023fabrication, currie1995nomenclature}. The LOD was calculated using the following equation: $\text{LOD} = \frac{3 \text{RMS$_{noise}$}}{S}$. The slope (S) has been calculated from the linear fitting of response vs concentration (ppb), as shown in Figure~\ref{fig:S3}. To calculate the $\text{RMS}_{noise}$, we first took 399 consecutive data points from the baseline response (\%) when NO$_2$ gas was turned off for each device. Finally, the data points were fitted using fifth order of polynomial to obtain the residual sum of squared (RSS). The fifth order polynomial fitting is shown in Figure~\ref{fig:S4}. Finally, the RMS$_{noise}$ can be calculated using: $\text{RMS}_{noise} = \sqrt{\frac{RSS=\sum^{N}_{i=1}{(\text{R}_i - \bar{\text{R}})^2}}{N}}$. 

\newpage


\newpage
\begin{figure}[ht!]
    \centering
    \includegraphics[width=0.99\linewidth]
    {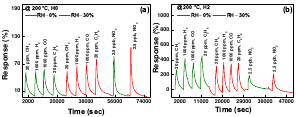}
    \caption{Selectivity measurements for different gases at 200 $^{\circ}$C at 0\% and 30\% humidity. (a) Unpatterned device (H0). (b) Honeycomb MoS$_{2}$ nanomeshes of initial circular etching pits R--250 nm (H2).}
    \label{fig:S5}
\end{figure}

\newpage

\section{NO$_2$ sensing at room temperature in the dark: an improved baseline recovery at high humidity levels}

\begin{figure}[ht!]
    \centering
    \includegraphics[width=1.0\linewidth]
    {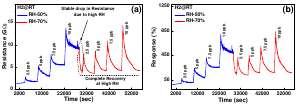}
    \caption{Room temperature NO$_{2}$ sensing response (2.5 -- 10 ppb) in the dark obtained from the honeycomb MoS$_2$ nanomesh device (H2) under 50\% (blue) and 70\% (red) humidity. (a) Resistance \textit{vs} time, (b) Sensing response (\%) \textit{vs.} time.}
    \label{fig:S9}
\end{figure}

Further, we evaluated the NO$_{2}$ at RT across different humidity levels in the dark. It is well-known that humidity poses a challenge for many gas sensor platforms. Regrettably, MoS$_2$-based sensors have yet to demonstrate satisfactory NO$_{2}$ sensing capabilities in high humidity environments at room temperature. Thus, it is an open problem to develop an NO$_{2}$ sensor with excellent humidity-tolerant performance~\cite{xia2019sulfur, agrawal2021strategy}. We measured NO$_{2}$ sensing at RT in the dark for 50\% and 70\% RH. The results for the honeycomb device H2 are displayed in Figure~\ref{fig:S9}. The blue curve is for 50\% RH, while the red curve is for 70\% RH. The increase in humidity is characterized by a sudden drop in the resistance as RH changes from 50\% to 70\%. It is worth noting that we have achieved exceptionally high responses of 83\% and 300\% for 2.5 ppb NO$_2$ at 50\% and 70\% humidity, respectively. These results, namely that our honeycomb nanomesh device H2 not only showed \textit{humidity-tolerant} behavior (200$^{\circ}$C in the dark) but also the \textit{enhanced} NO$_{2}$ sensing was activated by humidity at RT in the dark, are somewhat surprising. We also observed that the recovery to baseline resistance is improved considerably with an increase in RH. The honeycomb nanomesh device H2 has partially recovered at 50\% and completely recovered at high 70\% humidity for all NO$_{2}$ concentrations.

\newpage
\begin{figure}[ht!]
    \centering
    \includegraphics[width=0.99\linewidth]
    {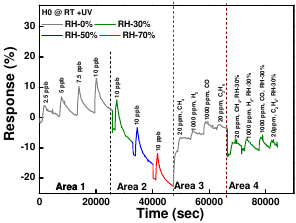}
    \caption{Broad NO$_{2}$ sensing response and selectivity measurements with and without humidity for unpatterned device (H0). Area 1: the first four cycles under 0\% humidity with UV illumination. Area 2: green (5$^{th}$), blue (6$^{th}$), and red (7$^{th}$) cycle for 10 ppb NO$_{2}$ at 30\%, 50\% and 70\% humidity, respectively. Area 3: selectivity measurements at 0\% humidity (8$^{th}$ -- 11$^{th}$ cycle). Area 4: selectivity measurements at 30\% humidity (12$^{th}$ -- 15$^{th}$ cycle).}
    \label{fig:S6}
\end{figure}

\newpage
\section{DFT relaxation}
\begin{figure}[ht]
    \centering
    \includegraphics[width=1.0\linewidth]{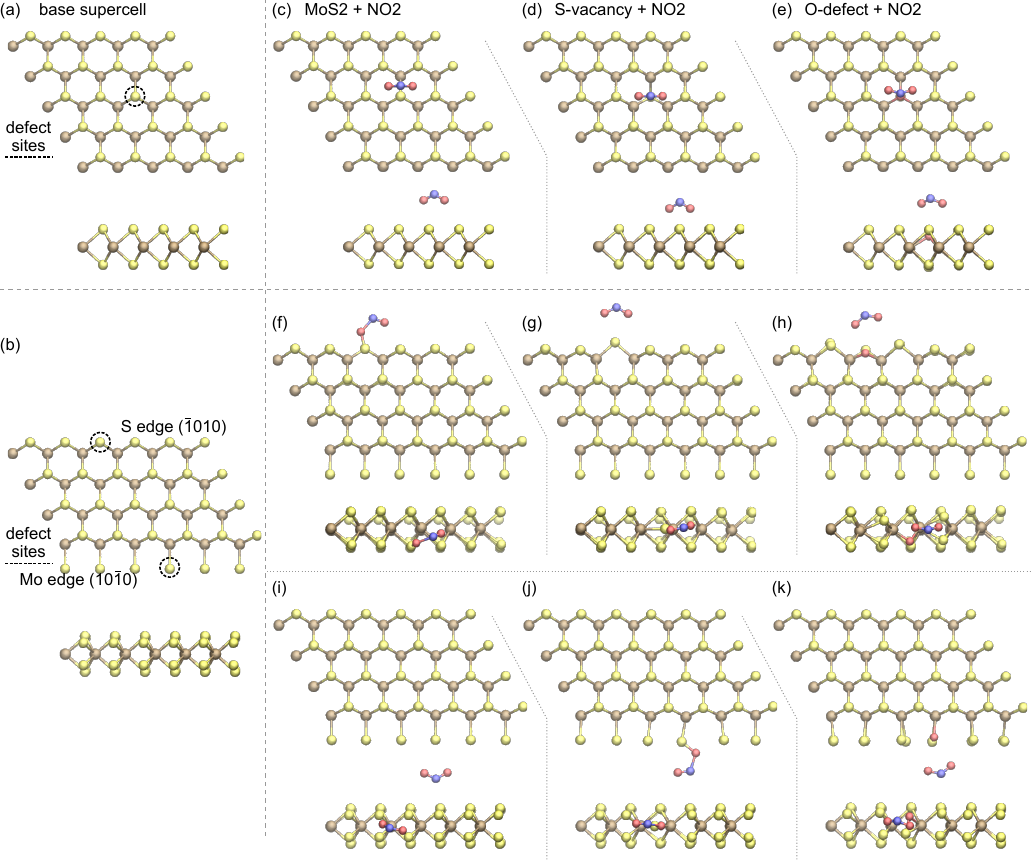}
    \caption{Results of ionic relaxation of the considered MoS$_2$ structures, including defects, with adsorbed NO$_2$.}
    \label{fig:Sdft}
\end{figure}

\newpage

\begin{figure}[ht!]
    \centering
    \includegraphics[width=0.99\linewidth]
    {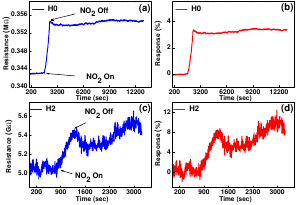}
    \caption{Room temperature NO$_{2}$ sensing response at 2.5 ppb in the dark (NO$_2$ is injected at points of rapid increase in the resistance). (a, b) Unpatterned device (H0). (c, d) Honeycomb MoS$_{2}$ nanomeshes of initial circular etching pits R--250 nm (H2). In both cases, the response does not return to the initial baseline level after the removal of NO$_2$.}
    \label{fig:SRT}
\end{figure}

\newpage
\begin{figure}[ht!]
    \centering
    \includegraphics[width=0.99\linewidth]
    {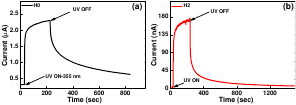}
    \caption{The photo-response curves obtained under UV illumination of (a) Unpatterned device (H0) and (b) Honeycomb MoS$_{2}$ nanomeshes of initial circular etching pits R--250 nm (H2).}
    \label{fig:S7}
\end{figure}

\newpage
\section{TableS1}
\begin{tabular}{|c|l|c|c|c|l|} 
        \hline 
        S.No & Sensing Materials & NO$_2$ conc. (ppb) & Temp. ($^{\circ}$C) & Response (\%) & Ref \\
        \hline
        1 & Ag-WO$_{3}$ & 100 & 200 & 2.6 & \cite{kamble2019ag} \\
         \hline
        2 & Au-In$_{2}$O$_{3}$ & 50 & 100 & 10 & \cite{zhang2023ppb} \\
         \hline
        3 & Co-SnS$_{2}$ & 1 & 190 & 89.7 & \cite{chang2023abundant} \\
         \hline
        4 & Au-CO$_{3}$O$_{4}$ & 10  & 136  & 5.80 & \cite{hsueh2021highly} \\
 \hline
        5 & Au@ZnO/rGO & 5 & 60 & 4.8& \cite{cao2022zno} \\
 \hline
        6 & ZnO nanorods-Pd & 100 & RT+VIS & 160 & \cite{wang2020highly} \\
         \hline     
        7 & MoS$_{2}$-ZnO & 50 & 60 & 63 & \cite{li2020nanocomposite} \\
         \hline
        8& ZnO-Ti$_{3}$C$_{2}$ & 10 & RT & 3 & \cite{liu2023zno} \\
         \hline
        9 & CaCl$_{2}$-Hydrogel    &80 & RT & 0.63& \cite{wu2021ion} \\
 \hline
        
10 & ZnO-Ti$_{3}$C$_{2}$ &10 & RT & 3&  \cite{liu2023zno}\\    
 \hline
11 & MoS$_{2}$-SnS    &25  &RT & 2.8&  \cite{kumar2022revealing}  \\
 \hline
12 & SnO$_{x}$-SnS & 1 & 350 & 170 & \cite{tang2021room} \\
 \hline
13 & MoS$_{2}$ @ SnO$_{2}$ & 10  & RT & 2 & \cite{Bai2021thin} \\
 \hline
14  & Cu$_{2}$O/CuO@rGO &30 & RT & 100& \cite{li2023one} \\
 \hline
15 & C-SnO$_{2}$ & 10 & RT & 11.4 & \cite{wu2021carbon} \\
 \hline
16 & SnS$_{2}$ QDs/rGO   & 1 & RT+VIS & 4 & \cite{huang2020sns2} \\
 \hline
17 & 2D/0D MoS$_{2}$/SnS & 25 & RT+UV & 0.26 & \cite{kumar2022revealing}\\
 \hline
18 & Porous ZnO/rGO &20 & RT & 2.38& \cite{chen2021porous} \\
 \hline
19 & 3D-Gr/CNTs &10 & RT & 9.1& \cite{hong2021enhanced} \\
 \hline
20 & MoS$_{2}$-GaSe& 20  & RT+VIS  & 6.3   & \cite{niu2021photovoltaic} \\
 \hline
21& InGaN & 200 & RT+UV & 6.12 & \cite{thota2021feather} \\
         \hline

22 & Hierarchical C-MoS$_{2}$ & 5 & RT & 167 & \cite{song2023mocvd} \\
 \hline
23 & BP & 0.4 & RT  & 8.9 & \cite{han2022sub} \\
 \hline
24 & MoS$_{2}$& 120  & RT+VIS & 50& \cite{cho2015bifunctional} \\
 \hline
25 & MoS$_{2}$ & 20 & RT & 21 & \cite{liu2014high} \\
         \hline
26 & BP Nanosheets & 50 & RT & 33 & \cite{ren2020improving} \\
         \hline
27 & Honeycomb MoS$_{2}$& 2.5& 200 & 100 & This Work\\
 \hline
28 & Honeycomb MoS$_{2}$ & 2.5 & RT+UV+RH-70\% & 1170 & This Work\\
 \hline
    \end{tabular}
    \label{table:S1}
\end{table}

\newpage
\section{Sensing mechanism}

In the following subsections, we propose the NO$_2$ sensing mechanism under various conditions such as NO$_2$ sensing in air, NO$_2$ sensing at high temperature, NO$_2$ sensing at RT under UV, NO$_2$ sensing at RT in dark with humidity and finally NO$_2$ sensing at RT in dark with variable humidity.

\subsubsection{Oxygen-induced doping of the nanopatterned MoS$_2$}
In air, MoS$_2$ adsorbs oxygen molecules and chemically interacts with them~\cite{Jaegermann1986,Qiu2012}. The local surface geometry and stoichiometry play a paramount role in MoS$_2$ reactivity. For instance, the step and edge sites were verified to be responsible for strong interactions with ambient O$_{2}$ both theoretically and experimentally~\cite{Jaegermann1986,Longo2017,Martincova2017}. Thus, such interactions are particularly feasible on the numerous edges of the nanofabricated MoS$_2$ nanomeshes. Sulfur vacancies are also responsible for the strong binding of oxygen. For instance, Nan \textit{et al.} reported that O$_{2}$ could be chemically adsorbed to monosulfur vacancies of defected monolayer MoS$_2$ with the strong binding energy of 2.295 eV (in the range of covalent bonding) and charge transfer value of 0.997$e$. In turn, at the perfect MoS$_2$ surface, the binding energy and charge transfer were just 0.012 eV and 0.021$e$, respectively~\cite{nan2014strong}. The interaction with oxygen proceeds beyond chemisorption: \textit{via} the dissociation of oxygen molecules towards the formation of oxidized Mo and S species on the surface~\cite{Martincova2017,Afanasiev2019}. For instance, progressive oxidation of the edge sites may lead to the local accumulation of Mo$^{+6}$ by-products~\cite{Jaegermann1986}. It was also calculated that nanoscale patches or chain-like structures can be formed along oxidized MoS$_2$ edges~\cite{Martincova2017,Szoszkiewicz2021}. STM visualization of sulfur vacancies saturated with the oxygen atoms upon ambient exposure was reported by Pető \textit{et al.}~\cite{Peto2018}. Therefore, it is highly likely that the edge sites and sulfur vacancies in our honeycomb MoS$_2$ nanomeshes are capped by oxygen. 
Oxidative wet etching employed in the deterministic fabrication of nanomeshes can also cause the formation of such patches. Interestingly, a distinct oxygen signal was found by electron energy loss spectroscopy (STEM-EELS) along zigzag edges of WS$_2$ nanomeshes fabricated using an identical methodology~\cite{munkhbat2020transition}. Here, we would like to point out that the capping of sulfur vacancies and edge sites with oxygen atoms should cause a significant $p$-doping of our MoS$_2$ sensing channels. Although some oxygen-capped sulfur vacancies are also expected in the non-patterned device H0, numerous zigzag edges of honeycomb nanomeshes are exceptionally favorable sites for introducing $p$-doping by oxygen. The doping is manifested in the \textit{I-V} characteristics of the fabricated devices {(Figure S2)}. The unpatterned device H0 and nanomesh H1 (with quite "diluted" patterning) exhibit hundreds-nA current at 1 V of applied bias, while the current through the denser nanopatterned nanomeshes H2 and H3 drops to the few-nA range. Note that this drastic change can not be explained just by subtraction of MoS$_2$ materials from the conductive channel due to the nanopatterning.

\subsubsection{NO$_2$ response at room temperature in the dark}
In dry air, the surface of fabricated sensing channels adsorbs molecular oxygen. Regardless of the particular adsorption site, O$_2$ molecule adsorption can be described by the following equations:

\begin{reaction}
  O$_2$ (gas) -> O$_{2}$ (ads)(physisorption)
\end{reaction}

\begin{reaction}
  O$_2$ (gas) + e^{-} -> O$_{2}^{-}$ (ads)
\end{reaction}

The latter chemisorption reaction assumes that O$_2$$^-$ ions are the most stable below 100$^{\circ}$C~\cite{Shim2018,bisht2022tailoring}. The oxygen-covered surface is unfavorable for the adsorption of other gases. Nowadays, there is insufficient knowledge about NO$_2$ adsorption on the MoS$_2$ with the defects capped by covalently bound oxygen atoms. 
 
In the recent two decades, NO$_2$ adsorption on the sensing and catalytic materials was increasingly studied by X-ray photoelectron spectroscopy (XPS), particularly in the N\textit {1s} region. Typically, such studies involve the materials' interaction with NO$_2$ in the reaction chamber and then the XPS measurement in an ultra-high vacuum (UHV) chamber~\cite{Baltrusaitis2009}. Therefore, the NO$_2$ adsorption fingerprints can be significantly affected by UHV conditions. Ambient pressure XPS (APXPS) can give more reliable data, but to the best of our knowledge, the available APXPS reports are limited to NO$_2$ adsorption on some of the metal oxides~\cite{Rosseler2013,Karagoz2021Cu2O,Karagoz2021CuOMoO3}. According to the reported UHV XPS data, N\textit{1s} region of MoS$_2$-based sensing materials exposed to NO$_2$ contains a broad signal, which is usually fitted by a stronger peak centered at 405.5-406.4 eV~\cite{Ikram20197,Ikram201911,Bai2021thin,Liu2020,Liu2023} and a smaller shouldered peak at 402.55-403.6 eV~\cite{Ikram20197,Ikram201911,Bai2021thin,Liu2020}. While a few authors assign the first peak to nitrate ions, its position does not correspond to the binding energies observed in NaNO$_3$ reference compound or adsorbed NO$_3$$^-$ on the surface of metal oxides (407.0-408.5 eV)~\cite{Ozensoy2005,Baltrusaitis2009,Rosseler2013,Nanayakkara2013}. Instead, it coincides with the N\textit{1s} binding energy of physisorbed NO$_2$/N$_2$O$_4$~\cite{Rosseler2013,Ozensoy2005,Baltrusaitis2009}. TPD of NO$_2$ from MoS$_2$/WS$_2$ surface confirmed a large fraction of weakly bound NO$_2$ desorbing below 200$^{\circ}$C~\cite{Ikram20197}. The second peak (402.55-403.6 eV) corresponds well to the nitrite signal~\cite{Rosseler2013,Baltrusaitis2009,Ozensoy2005,Ikram20197,Ikram201911,Bai2021thin}. Thus, the NO$_2$ adsorption on MoS$_2$ at room temperature can be described by the following equations:

\begin{reaction}
  NO2 (gas) -> NO_{2}(ads)
\end{reaction}
\begin{reaction}
  NO2 (gas) + e^{-} -> NO_{2}^{-}(ads)
\end{reaction}

Nevertheless, we still could detect a few-percent change in the resistivity of both non-patterned device H0 and nanomesh device H2 upon introducing 2.5 ppb NO$_2$ at RT in the dark (Figure~\ref{fig:SRT}). However, this response was irreversible due to strong binding of NO$_2$ on MoS$_2$ surface. The low NO$_2$ response at RT and in dark is due to less availability of NO$_2$ adsorption sites which are already occupied with oxygen. Similar results have also been verified in our DFT calculations. At room temperature and in the dark, oxygen removal from MoS$_2$ surface is suppressed, as demonstrated using temperature-programmed desorption (O$_2$-TPD) by Ikram \textit{et al.}~\cite{Ikram201911}. Hence, without any additional energy input activating surface chemical reactions, the NO$_{2}$ response of our MoS$_2$-based devices was low Figure~\ref{fig:SRT}.

\subsubsection{UV illumination enables high NO$_2$ sensitivity at room temperature}

Although RT operation of MoS$_2$-based NO$_2$ sensors is complicated for the reasons mentioned above, it is still extremely attractive because of better sensor lifespan, low power consumption, and compatibility with wearable electronics and flammable gas environments. One of the feasible strategies enabling RT sensitivity to NO$_2$ is the illumination of sensitive MoS$_2$ layer with UV light~\cite{agrawal2021strategy}. Here, we used UV-LED with a center wavelength of 355 nm. UV light activates several physical and chemical processes in our honeycomb nanomesh devices, including the generation of electron-hole pairs in MoS$_2$, accelerated adsorption and desorption of gas molecules, and photochemical reactions in the adsorbed layers.
 
Upon UV illumination of the MoS$_{2}$ surface, an extensive photogeneration of electron-hole (\textit{e-h}) pairs occurs. This effect is significantly more pronounced in honeycomb MoS$_{2}$ nanomeshes than in the unpatterned device, H0. The current \textit{vs.} time analysis, performed under UV exposure for both devices (Figure S7), reveals that the honeycomb MoS$_{2}$ nanomeshes H2 not only exhibit a considerably higher photocurrent relative to their dark current but also demonstrate a shorter decay time in comparison to the unpatterned device H0. The role of these photogenerated \textit{e-h} pairs is critical for NO$_{2}$ detection. The photogenerated holes $h^{+}_{(h\nu)}$ can recombine with the electrons associated with both O$_{2}^{-}$ (ads) and NO$_2$$^-$ (ads) facilitating their desorption from MoS$_2$ surface. Simultaneously, the surplus of photogenerated electrons $e^{-}_{(h\nu)}$ provides more opportunities for chemisorption of O$_2$ and NO$_2$ molecules. The net thermodynamic effect can be complex. For oxygen alone, the net mass balance on the surface seems to depend on the O$_2$ partial pressure, UV wavelength (in particular, its ability to generate ozone from oxygen), and other factors. For instance, Chee \textit{et al.} reported gradual accumulation of the oxidated species upon UV illumination of monolayer MoS$_2$ in 99\%/1\% N$_2$/O$_2$ atmosphere (RT, UV lamp with 253.7 nm (90\%) and 184.9 nm (10\%) wavelengths)~\cite{Chee2020}. In turn, Wang \textit{et al.} reported oxygen photodesorption from monolayer MoS$_2$ under UV illumination in vacuum (RT, UV LED with 397 nm center wavelength)~\cite{Wang2020UV}. Note that we used UV LED with a 355 nm center wavelength, which cannot generate ozone from molecular oxygen (without additional mediators).

The boosted adsorption and desorption processes make the conditions on the UV-illuminated MoS$_2$ surface much more "dynamic" (and, potentially, far from equilibrium) than in the dark. Hence, NO$_2$ and O$_2$ competitively adsorb on / desorb from the MoS$_2$ nanomeshes according to the following equations:

\begin{reaction}
  O2^{-} (ads) + h^{+}_{($h\nu$)}  -> O2 (gas)
\end{reaction}
\begin{reaction}
  O2 (gas) + e^{-}_{($h\nu$)}  -> O2^{-} (ads)
\end{reaction}
\begin{reaction}
  NO2 (gas) + e^{-}_{($h\nu$)} -> NO_{2}^{-} (ads)
\end{reaction}
\begin{reaction}
  NO2^{-} (ads) + h^{+}_{($h\nu$)} -> NO2 (gas)
\end{reaction}

In such a dynamic environment enabled by photogenerated e-h pairs, higher electron affinity makes even single-ppb concentrations of NO$_2$ detectable in the presence of 200000 ppm background O$_2$ and the initial four sensing cycles shown in Figure 6b of the main text. The UV light also reduces the energy barrier for NO$_{2}$ desorption~\cite{zheng2020mos2}. Hence, the full recovery is obtained under UV light illumination of TMD sensors as demonstrated in literature~\cite{azizi2020high} and our gas sensing experiments.

\subsubsection{Enhanced NO$_2$ sensitivity at 200$^{\circ}$C in the dark}

As the second approach to surmounting the low NO$_2$ response at RT, we used heating MoS$_2$ nanomeshes to the elevated temperature of 200$^{\circ}$C. O$_2$ TPD data reported elsewhere~\cite{Ikram201911} shows a strong peak at 185$^{\circ}$C indicating the removal of labile oxygen species (presumably physisorbed or weakly chemically bonded to MoS$_2$ surface). Thus, at 200$^{\circ}$C, the kinetics of sorption-desorption processes at the MoS$_2$ surface can be efficiently accelerated, resembling the case of UV illumination. Furthermore, the elevated temperature seems to have a clearer thermodynamic effect: thermal energy ($k_{B}T$) makes the O$_2$ desorption more favorable than NO$_2$ desorption because NO$_2$ has higher binding energy to MoS$_2$ (and electron affinity) than O$_2$. Thereby, elevated temperature expedites the removal of oxygen adsorbed on the surface and creates more fresh available sites for NO$_2$ adsorption. Additionally, upon increasing the temperature, chemisorbed O$_2$$^-$ ions may convert into O$^-$, which is a more stable between 100$^{\circ}$C and 300$^{\circ}$C~\cite{Shim2018,bisht2022tailoring}:

\begin{reaction}
  O$_{2}^{-}$ (ads) + e^{-} -> 2 O$^{-}$ (ads)
\end{reaction}
The chemisorbed NO$_x$ species can also be transformed upon elevating temperature. In the case of $\theta$-Al$_2$O$_3$ films, the heating from 80 K to RT resulted in the conversion of adsorbed molecular NO$_2$/N$_2$O$_4$ into NO$_3$$^-$~\cite{Ozensoy2005}. This process seems to involve NO$_2$ disproportionation with subsequent NO desorption:
\begin{reaction}
  2 NO$_{2}$ (ads) -> NO$^{+}$ (ads) + NO$_3$$^-$ (ads)
\end{reaction}
\begin{reaction}
  NO$^{+}$ (ads) + e- -> NO (gas)
\end{reaction}

By analogy, the thermal activation might result in the formation of surface nitrates upon heating MoS$_2$ nanomeshes from RT to 200$^{\circ}$C. O$^{-}$ ions can also contribute to nitrate formation according to the following equation:

\begin{reaction}
  NO2 (gas) + O$^{-}$ (ads) -> NO$_{3}^{-}$ (ads)
\end{reaction}

However, the formation of nitrates on the MoS$_2$ surface still requires experimental validation.
The superior NO$_{2}$ sensing performance of honeycomb MoS$_{2}$ nanomeshes H1-H3 relative to their unpatterned counterpart H0 (Figure 3a-d) can be elucidated by examining the unique structural and chemical properties that enhance NO$_{2}$ adsorption. The edges of the hexagonal pits in MoS$_{2}$ nanomeshes offer numerous sites of high chemical reactivity. The atomic structure at these edges provides favorable conditions for the adsorption of NO$_{2}$, including increased binding energy and charge transfer compared to the less reactive basal plane. These beneficial features can lead to a stronger interaction between the NO$_{2}$ molecules and the MoS$_{2}$ surface, facilitating more efficient adsorption and charge transfer. Moreover, at the elevated operating temperature, the O$_2$ desorption is higher at the edges of the hexagonal pits and leads to increased availability of fresh active sites on the MoS$_{2}$ nanomeshes. This continuous availability of fresh active sites ensures sustained sensing performance over time for honeycomb MoS$_{2}$ nanomeshes. Concurrently, increased temperature leads to enhanced thermal oscillations of NO$_{2}$ molecules and a higher desorption rate that surpasses the adsorption rate on MoS$_{2}$~\cite{cho2015highly}. Thereby, a complete recovery of the adsorption sites is observed at 200$^{\circ}$C, in contrast to RT. This phenomenon of heightened NO$_{2}$ sensitivity and comprehensive recovery is delineated in the sensing profiles presented in Figure 3a-d, with comparative data at RT provided in the Supplementary Information, Figure S8.


\subsubsection{Humidity-tolerant behavior and humidity-enhanced NO$_{2}$ sensitivity at 200$^{\circ}$C in the dark}

We have introduced humidity in our sensing measurements at three different operating conditions: (1) high temperature of 200$^{\circ}$C in the dark, (2) RT in the dark, and (3) RT with UV illumination. In all of these conditions, the NO$_{2}$ sensing was enhanced with the increasing humidity from 0\% to 70\%. Interestingly, the best response values were observed under UV illumination with the highest humidity of 70\%, which is a strongly deteriorating condition for many reported gas sensors~\cite{wang2022recent}. Furthermore as shown in Figure 4a-b, at 200$^{\circ}$C operating temperature, the increase of the NO$_2$ response was very small in the 30\%-70\% RH range, \textit{i.e.} our nanofabricated devices demonstrated a stable performance and anti-humidity behavior in the typical conditions of indoor and outdoor air quality monitoring.

The effects of humidity on MoS$_2$ properties have been extensively studied for decades since the utilization of this material as a solid lubricant began~\cite{bobbitt2022interactions}. The widely accepted mechanism of water adsorption on the MoS$_2$ surface has three distinct stages: (\textbf{I}) chemisorption, (\textbf{II}) physisorption of 1-2 partially immobilized layers, and (\textbf{III}) multilayer physisorption of labile quasi-liquid H$_2$O film~\cite{siddiqui2022highly, Wang2021PEO,lei2016first}.
 
Stage \textbf{I}. First, the water molecules are adsorbed on the MoS$_2$ surface. Different theoretical reports predict that the defect-free MoS$_2$ basal plane either repulse~\cite{Ghuman2015} or slightly attracts~\cite{bobbitt2022interactions} water molecules. The latter prediction agrees with the slight hydrophilicity (69$^{\circ}$ water contact angle) of freshly cleaved MoS$_2$ basal surface~\cite{Kozbial2015}. In turn, sulfur vacancies and O-substitution defects are predicted to strongly adsorb water~\cite{bobbitt2022interactions,Ghuman2015}. Furthermore, water molecules can spontaneously dissociate on the MoS$_2$ defects, and (particularly relevant for our MoS$_2$ nanomeshes) Mo-terminated and under-coordinated S-terminated edges are the preferable sites for this process~\cite{Ghuman2015}. Dissociative chemisorption results in a layer of OH-groups bound to the MoS$_2$ surface, while H$^+$ is more labile~\cite{Ghuman2015,siddiqui2022highly,wang2022recent,lei2016first}. The as-formed layer of chemisorbed OH-groups acts as a scaffold for the subsequent water physisorption.
 
Stage \textbf{II}. With increasing relative humidity, more water molecules approach the MoS$_2$ surface and form the first physisorbed layer bound to the chemisorbed OH-groups by hydrogen bonds. 
In defect-rich MoS$_2$ nanomeshes, the surface OH-groups can be arranged densely enough to form two hydrogen bonds with each water molecule of the first physisorbed layer and significantly restrict both their own movement and proton transport capability~\cite{zhen2014humidity}. Starting from the second physisorbed layer, H$_2$O molecules become more and more labile~\cite{zhen2014humidity,bi2013ultrahigh}.

Stage \textbf{III}. At high RH, multilayer physisorption results in a water film with a quasi-liquid behavior and plenty of dynamic hydrogen bonds formed between the mobile H$_2$O molecules. In this medium, self-ionization of water occurs according to the following equation:

\begin{reaction}
  2 H2O  -> H3O$^+$ + OH$^-$
\end{reaction}

Once the bias is applied to the MoS$_2$ sensing channel paced into a high RH environment, the ionic current can easily flow in the quasi-liquid water film \textit{via} Grotthuss mechanism, \textit{i.e.} proton hopping between the clusters of H$_2$O molecules~\cite{agmon1995grotthuss}. Protons formed during the dissociative water chemisorption on the edges and point defects of MoS$_2$ can also participate in this process. Potentially, the applied bias can also enhance the self-ionization of water.

Three distinct stages of water adsorption are clearly manifested in different characteristics of MoS$_2$-based devices, including saturation resistance and dynamic resistive response at 10\%--80\% RH~\cite{Wang2021PEO}, hysteresis of humidity-related capacitive response at 11\%--95\% RH~\cite{lei2016first}, and surface potential \textit{in situ} visualized using Kelvin probe force microscopy at 5\%--95\% RH~\cite{Feng2017}. From these reports, it can be estimated that the water chemisorption dominates till 40\%--54\% RH (transition between the stages \textbf{I} and \textbf{II}), and the quasi-liquid water layer is finalized at 60\%--75\% RH (transition between the stages \textbf{II} and \textbf{III}). A certain variation of these values seems to be due to the unequal defect concentrations in the utilized MoS$_2$ devices and the different nature of the registered responses.
 
Water adsorption upon increasing RH also affected the NO$_2$ sensing response of MoS$_2$ nanomeshes developed in this work. At 200$^{\circ}$C in the dark, the background resistance sharply increased upon switching between 0\% RH and 30\% RH as shown in Figure 4a-b. Also, the response to 10 ppb NO$_2$ was significantly higher at 30\% RH than in dry synthetic air (Figure 4c). This implies that at 200$^{\circ}$C in the synthetic air, water chemisorption extracts electrons from the unpatterned device H0 and honeycomb MoS$_{2}$ nanomesh device H2:

\begin{reaction}
   H2O (gas) + O$^-$ (ads) + e$^-$ -> 2 OH$^-$ (ads)
\end{reaction}

The change in resistance in device H0 is lower than in H2. The honeycomb MoS$_{2}$ nanomesh has a much higher number of oxidized edge sites along the hexagonal etching pits, which can chemisorb more water molecules than the unpatterned basal MoS$_2$ surface. However, upon further increase in humidity up to 70\% RH, no changes in the background resistance and just marginal growth of NO$_2$ response were observed in both devices (Figure 4a-b). Obviously, at high temperatures, only the stage \textbf{I} of water adsorption takes place where H$_2$O molecules dissociate and form a layer of OH-groups on the MoS$_{2}$ surface. The further water physisorption (stages \textbf{II} and \textbf{III}) is suppressed due to high temperature~\cite{kwak2018humidity}.
 
In turn, at room temperature in the dark, we detected a significant drop in the resistance of H2 device upon switching from 50\% RH to 70\% RH (Figure 5). This observation is similar to the rapid drop in the resistance of liquid-phase exfoliated MoS$_2$ nanosheets in the 40\%--60\% RH range reported by Wang \textit{et al.}~\cite{Wang2021PEO}. In both cases, the resistance drop is due to the appearance of physisorbed layers of increasingly mobile H$_2$O molecules, resulting in more and more facilitated proton hopping and enhanced ionic conductivity in the overall inter-electrode channel~\cite{Wang2021PEO,wang2022recent,zhen2014humidity}.

In the context of humidity-activated NO$_{2}$ response, the NO$_{2}$ sensing behavior is notably influenced by the presence of water or through its reactions with surface-bound hydroxyl ions (OH$^{-}$) and protons (H$^{+}$). These interactions can lead to the creation of surface-attached nitrate ions (NO$_{3}^{-}$(ads)), which are marked by their increased binding energy and a more pronounced transfer of charge on MoS$_{2}$, thereby amplifying the NO$_{2}$ detection capabilities. Yao \textit{et al.} reported the development of a humidity-assisted NO$_{2}$ sensor utilizing carbon nanotubes (CNTs) as the sensing material~\cite{yao2011humidity}. In their study, the adsorption of NO$_{2}$ on the CNT surface results in the formation of NO$_{3}^{-}$ ions. Furthermore, the computational analysis revealed that the binding energies for NO$_{2}^{-}$ and  NO$_{3}^{-}$ on CNT surfaces are 0.47 eV and 1.08 eV, respectively. This computational insight provides a basis for expecting similar adsorption dynamics in our investigations. Some similar findings were also observed in 2D materials based NO$_{2}$ sensors~\cite{bisht2022tailoring, han2019interface, song2019ionic}. Another significant reaction pathway involves NO$_{2}$ interaction with surface-adsorbed hydroxyl groups. Nevertheless, the efficiency of these reactions decreases on MoS$_{2}$ surface if that is heavily covered with water layers, as the NO$_{2}$  molecules encounter challenges in accessing the MoS$_{2}$  sensor surface, needing to traverse through the thick layers of water, which might reduce the sensor sensitivity. The honeycomb MoS$_{2}$ nanomeshes have shown humidity-activated NO$_{2}$ response in both dark and UV illumination. Our experimental findings under dark revealed that the honeycomb MoS$_{2}$ nanomeshes sensor response to NO$_{2}$ reached a point of saturation with slight improvement only (from 536\% to 661\%) when the relative humidity increased from 50\% to 70\% in the absence of light. This may be due to increased multilayer water formation on honeycomb MoS$_{2}$ nanomeshes surface in the dark. There may be the possibility that if we further increase the humidity beyond 70\%, the response may decrease. But, achieving such high humidity is beyond our setup limitation.

\subsubsection{Humidity activated faster recovery}
The enhancement of NO$_2$ recovery in our study is a significant observation, particularly under conditions of humidity, both in dark environments and under UV illumination~\cite{azizi2020high}. This phenomenon can be attributed to the polar nature of NO$_2$ and H$_2$O molecules, in contrast to the relatively nonpolar background gases such as N$_2$ and O$_2$. Due to their polar characteristics, H$_2$O molecules exhibit stronger dipole-dipole interactions with NO$_2$ compared to N$_2$ and O$_2$. Consequently, as H$_2$O molecules pass near NO$_2$ adsorbed on the MoS$_2$ surface, they effectively attract and remove the NO$_2$ molecules from the surface, thereby significantly enhancing the recovery rate under humid conditions. This effect becomes more pronounced with increasing humidity, as a greater volume of H$_2$O molecules facilitates the desorption of NO$_2$ from the MoS$_2$ surface. A similar phenomenon has also been observed in previously reported elsewhere~\cite{azizi2020high}.

\newpage
\printbibliography[heading=subbibintoc]